 \documentstyle[11pt]{article}

\font\fr=eufm10 scaled \magstep 1 
\font\es=msbm10                   
\newtheorem{teor}{Theorem}
\newtheorem{prop}{Proposition}

\newtheorem{definition}{Definition}
\newtheorem{lem}{Lemma}
\def\beq{\begin{equation}}
\def\eeq{\end{equation}}
\def\bea{\begin{eqnarray}}
\def\eea{\end{eqnarray}}
\def\beann{\begin{eqnarray*}}
\def\eeann{\end{eqnarray*}}
\def\ben{\begin{enumerate}}
\def\een{\end{enumerate}}
\def\bit{\begin{itemize}}
\def\eit{\end{itemize}}
\def\dst{\(\displaystyle}
\def\derpar#1#2{\frac{\partial{#1}}{\partial{#2}}}

\def\feble#1{\mathrel{\mathop =\limits_{#1}}}

\def\moment#1#2#3{{#1}_{#2}, \ldots, {#1}_{#3}}

\def\qed{\ifvmode\removelastskip\fi
{\unskip\nobreak\hfil\penalty50\hbox{}\nobreak\hfil \hbox{\vrule
height1.2ex width1.2ex}\parfillskip=0pt \finalhyphendemerits=0
\par\smallskip}}
\def\vf{\mbox{\fr X}}
\def\df{{\mit\Omega}}
\def\Lag{{\cal L}}
\def\lag{\pounds}

\def\d{{\rm d}}

\def\Real{\mbox{\es R}}

\def\inn{\mathop{i}\nolimits}
\def\Tan{{\rm T}}

\def\Lie{\mathop{\rm L}\nolimits}
\def\ls{((E,M;\pi),\Lag )}
\def\hs{(J^{1*}E,\Omega^\nabla_h)}
\def\hso{(J^{1*}E,P,\Omega^0_h)}
\def\hsl{(J^{1*}E,\Omega_h)}
\def\del{{\cal E}^{\nabla}_{\Lag}}

\def\Cinfty{{\rm C}^\infty}
\def\proof{( {\sl Proof} )\quad}

\def\tabaddress#1{{\small\it\begin{tabular}[t]{c}#1 \\[1.2ex]\end{tabular}}}

\title{MULTIVECTOR FIELD FORMULATION OF HAMILTONIAN FIELD THEORIES:
       EQUATIONS AND SYMMETRIES}
\author{\sc A. Echeverr\'{\i}a-Enr\'\i quez,
M.C. Mu\~noz-Lecanda
\thanks{{\bf e}-{\it mail}: MATMCML@MAT.UPC.ES},
N. Rom\'an-Roy
\thanks{{\bf e}-{\it mail}: MATNRR@MAT.UPC.ES}
   \\
  \tabaddress{\it Departamento de Matem\'atica Aplicada y
  Telem\'atica.\\
 Edificio C-3, Campus Norte UPC.  C/ Jordi Girona 1.
   E-08034 Barcelona. SPAIN}}
\date{(math-ph/9907007) }

\begin{document}

 \maketitle
 \thispagestyle{empty}
 \setcounter{page}{0}

 \begin{abstract}
 We state the intrinsic form of the Hamiltonian equations
 of first-order Classical Field theories
 in three equivalent geometrical ways: using
 multivector fields, jet fields and connections.
 Thus, these equations are given in a form
 similar to that in which the Hamiltonian equations of
 mechanics are usually given.

 Then, using multivector fields, we
 study several aspects of these equations,
 such as the existence and
 non-uniqueness of solutions, and the integrability problem.
 In particular, these problems are analyzed for the case
 of Hamiltonian systems defined in a submanifold of the
 multimomentum bundle. Furthermore, the existence
 of first integrals of these Hamiltonian equations is considered,
 and the relation between {\sl Cartan-Noether symmetries} and
 {\sl general symmetries} of the system
 is discussed. Noether's theorem is also stated in this
 context, both the ``classical'' version and its generalization
 to include higher-order Cartan-Noether symmetries.
 Finally, the equivalence between the Lagrangian
 and Hamiltonian formalisms is also discussed.
 \end{abstract}


 \bigskip  \bigskip
 AMS s.\,c.\,(1991): 53C80, 55R10, 58A20, 70G50, 70H99.
 \qquad
 PACS: 0240, 0320, 0350.

 \clearpage

 \section{Introduction}

 The geometric structures underlying the covariant
 Lagrangian description of first-order Field theories are first
 order jet bundles $J^1E\stackrel{\pi^1}{\to}E\stackrel{\pi}{\to}M$
 and their canonical structures (see \cite{EMR-96}, and
 references quoted therein). For the covariant Hamiltonian
 formalism several formulations arise, which use different
 kind of differentiable structures ({\sl polisymplectic,
 $k$-symplectic, $k$-cosymplectic} or {\sl multisymplectic} forms)
 and {\sl multimomentum phase spaces}
 where the formalism takes place
 (see, for instance, \cite{Aw-92}, \cite{CCI-91},
 \cite{GMS-97}, \cite{GIMMSY-mm}, \cite{Gu-87}, \cite{Ka-98},
 \cite{KT-79}, \cite{LMO-98}, \cite{Pu-88}, \cite{Sd-95}).

 In any case, a subject of interest in the geometrical description of the
 Hamiltonian formalism of Classical Field theories is related to
 the field equations, which are called the
 {\sl Hamiltonian equations}.
 In the multisymplectic models, both in the Lagrangian and Hamiltonian
 formalisms, the field equations are usually written using the
 multisymplectic form in order to characterize the critical
 sections which are solutions of the problem
 \cite{EMR-96}, \cite{Gc-73}, \cite{GS-73}.
 This characterization can be derived from a suitable
 variational principle.

 However, other attempts have been made to write these
 field equations in a more geometric-algebraic manner (as is
 done in mechanics, using vector fields); namely:
 by using {\sl  Ehresmann connections}
 \cite{LMM-95}, \cite{Sa-89}, {\sl jet fields} \cite{EMR-96},
 or {\sl multivector fields} \cite{GS-73}, \cite{Ka-93}, \cite{Ka-95},
 \cite{Ka-97b}, \cite{Ka-98}.
 All of them have been carefully studied in \cite{EMR-98} for the
 Lagrangian formalism of Field theories, and their equivalence
 demonstrated. The aim of this work
 is to carry out the analysis of these
 procedures for the Hamiltonian formalism, proving that all of them
 are equivalent, and using in particular the
 multivector field formulation to
 study the existence and non-unicity of solutions
 of these equations, and their integrability.
 Furthermore, equivalence theorems between the Lagrangian and
 Hamiltonian formalisms are stated.
 Thus, previous works of I.V. Kanatchikov devoted to the analysis of
 the field equations in the Hamiltonian formalism using multivector
 fields (in a more specific context), are completed.

 Another subject of interest is the study of
 symmetries. Again using the multivector field formalism,
 we introduce and characterize different kinds of symmetries
 which are relevant in Field theory, showing their relation.
 In particular, Noether's theorem is proved and
 generalized in order to include higher-order Noether symmetries.

 The paper is structured as follows:
 In Section 2 we review the construction of Hamiltonian systems in
 Field theory. Section 3 is devoted to setting the Hamiltonian field
 equations in terms of multivector fields, connections and jet
 fields (showing the equivalence of three methods), analyzing
 the existence and non-uniqueness of solutions (in the regular
 case), and their integrability. Sections 4 and 5 deal with the
 study of symmetries, first integrals and Noether's theorem.
 In Section 6, the case of restricted Hamiltonian systems is
 considered (those where the equations are defined in a submanifold
 of the multimomentum bundle). Hamiltonian systems associated with
 Lagrangian systems are treated in Section 7, including the equivalence
 between the Lagrangian and Hamiltonian formalism (for
 hyper-regular case). In Section 8, an example
 which is a quite general version of many typical models in Field theories
 is analyzed. The last Section is devoted to presenting the
 conclusions. The work ends with an appendix where the main
 features concerning multivector fields and connections are
 reviewed.

 All manifolds are real, paracompact, connected and $C^\infty$. All
 maps are $C^\infty$. Sum over crossed repeated indices is
 understood. Throughout this paper
 $\pi\colon E\to M$ will be a fiber bundle ($\dim\, M=m$, $\dim\, E=N+m$),
 where $M$ is an oriented manifold with volume form $\omega\in\df^m(M)$,
 and $\pi^1\colon J^1E\to E$ will be the
 jet bundle of local sections of $\pi$.
 The map $\bar\pi^1 = \pi \circ \pi^1\colon J^1E
 \longrightarrow M$ defines another structure of differentiable
 bundle. Finally, $(x^\mu,y^A,v^A_\mu)$ will be
 natural local systems of coordinates in $J^1E$
 ($\mu = 1,\ldots,m$; $A= 1,\ldots,N$).

 \section{Hamiltonian systems}

 The Hamiltonian formalism for first-order Field theories requires
 the choice of a multimomentum phase space.
 This choice is not unique.
 In \cite{EMR-99a} and \cite{EMR-99b}, the
 relations among some of them are shown, and in particular
 the following result is proved (see also \cite{CCI-91} and \cite{MS-99}):

 \begin{teor}
 Let $\pi\colon E\to M$ be a fiber bundle. Then the following
 bundles are diffeomorphic:
 \begin{enumerate}
  \item
 $\Lambda_1^m\Tan^*E/\pi^*\Lambda^m\Tan^*M$
 (where $\Lambda_1^m\Tan^*E\equiv {\cal M}\pi$
 is the bundle of $m$-forms on
 $E$ vanishing by the action of two $\pi$-vertical vector fields).
  \item
 ${\rm Aff}(J^1E,\pi^*\Lambda^m\Tan^*M)/\pi^*\Lambda^m\Tan^*M$
 (where ${\rm Aff}(J^1E,\pi^*\Lambda^m\Tan^*M)$ denotes the set
 of affine bundle maps from $J^1E$ to $\pi^*\Lambda^m\Tan^*M$).
  \item
  $\pi^*\Tan M\otimes{\rm V}^*(\pi)\otimes\pi^*\Lambda^m\Tan^* M$
 (where ${\rm V}^*(\pi)$ denotes the dual bundle of
 ${\rm  V}(\pi)=\ker\,\Tan\pi$).
 \end{enumerate}
 \end{teor}

 Thus, we take these equivalent bundles as our
 multimomentum phase space, and call it the
 {\sl multimomentum bundle}.
 We denote it by $J^{1*}E$, and its points as $\tilde y\in J^{1*}E$.
 For the natural projections we will write
 $\tau^1\colon J^{1*}E\to E$ and
 $\bar\tau^1=\pi\circ\tau^1\colon J^{1*}E\to M$.
 Given a system of coordinates adapted to the bundle
 $\pi\colon E\to M$, we can construct natural coordinates in
 $J^{1*}E$ and ${\cal M}\pi$, which will be denoted as
 $(x^\mu,y^A,p^\mu_A)$ and $(x^\mu,y^A,p^\mu_A,p)$, respectively.

 In order to complete the geometric background of the
 Hamiltonian formalism, the multimomentum bundle must be
 endowed with a geometric structure which characterizes the
 system. Thus, we can construct {\sl Hamiltonian systems}
 in three different ways
 \cite{CCI-91}, \cite{EMR-99b}, \cite{GMS-97}, \cite{LMM-96}, \cite{Sd-95}:

 First, the multicotangent bundle
 $\Lambda^m\Tan^*E$ is endowed with canonical forms \cite{CIL-98}:
 ${\bf \Theta}\in\df^m(\Lambda^m\Tan^*E)$
 and the multisymplectic form
 ${\bf  \Omega}:=-\d{\bf \Theta}\in\df^{m+1}(\Lambda^m\Tan^*E)$.
 But ${\cal M}\pi\equiv\Lambda^m_1\Tan^*E$ is a subbundle of
 $\Lambda^m\Tan^*E$. Then, if
 $\lambda\colon\Lambda^m_1\Tan^*E\hookrightarrow\Lambda^m\Tan^*E$
 is the natural imbedding,
 $ \Theta :=\lambda^* {\bf \Theta}$ and
 $\Omega :=-\d\Theta=\lambda^*{\bf \Omega}$
 are canonical forms in ${\cal M}\pi$, which are called
 the {\sl multimomentum Liouville} $m$ and $(m+1)$ {\sl forms}
 of ${\cal M}\pi$. In a system of natural coordinates in
 ${\cal M}\pi$ we have
 \beq
 \Theta = p_A^\mu\d y^A\wedge\d^{m-1}x_\mu+p\d^mx
 \quad , \quad
 \Omega = -\d p_A^\mu\wedge\d y^A\wedge\d^{m-1}x_\mu-\d p\wedge\d^mx
 \label{cero}
 \eeq

 A section $h\colon J^{1*}E\to {\cal M}\pi$ of
 the projection $\mu\colon{\cal M}\pi\to J^{1*}E$
 is called a {\sl Hamiltonian section}.
 The {\sl Hamilton-Cartan} $m$ and $(m+1)$ {\sl forms} associated
 with the Hamiltonian section $h$ are
 $$
 \Theta_h=h^*\Theta \quad ; \quad  \Omega_h=h^*\Omega=-\d\Theta_h
 $$
 Using natural coordinates in $J^{1*}E$,
 a Hamiltonian section is locally specified by a
 {\sl local Hamiltonian function}
 $H\in\Cinfty (U)$, $U\subset J^{1*}E$, such that
 $h(x^\mu,y^A,p^\mu_A)\equiv
 (x^\mu,y^A,p^\mu_A,p=-H(x^\gamma,y^B,p_B^\nu))$.
 Therefore,
 if $\bar\tau^{1*}\omega=\d^mx\equiv\d x^1\wedge\ldots\wedge\d x^m$,
 the Hamilton-Cartan forms take the local expressions
 \beq
 \Theta_h = p_A^\mu\d y^A\wedge\d^{m-1}x_\mu -H\d^mx
 \quad , \quad
 \Omega_h = -\d p_A^\mu\wedge\d y^A\wedge\d^{m-1}x_\mu +
 \d H\wedge\d^mx
 \label{omegaHlocal}
 \eeq
 where
 \dst\d^{m-1}x_\mu\equiv\inn\left(\derpar{}{x^\mu}\right)\d^mx\) .

 A variational problem can be posed for the system $(J^{1*}E,\Omega_h)$:
 the states of the field are the sections of
 $\bar\tau^1$ which are critical for the functional
 ${\bf H}\colon\Gamma_c(M,J^{1*}E)\to\Real$
 defined by ${\bf H}(\psi):=\int_M\psi^*\Theta_h$,
 for every $\psi\in\Gamma_c(M,J^{1*}E)$; where $\Gamma_c(M,J^{1*}E)$
 is the set of compact supported sections of $\bar\tau^1$.
 As is known \cite{EMR-96}, \cite{EMR-99b}, these
 critical sections are characterized by the condition
 $\psi^*\inn (X)\Omega_h=0$, for every $X\in\vf (J^{1*}E)$,
 which in natural coordinates in $J^{1*}E$,
 is equivalent to demanding that $\psi=(x^\mu,y^A(x),p^\mu_A(x))$
 satisfies the equations
 \beq
 \derpar{y^A}{x^\mu}\Big\vert_{\psi}=
 \derpar{H}{p^\mu_A}\Big\vert_{\psi}
 \quad ;\quad
 \derpar{p_A^\mu}{x^\mu}\Big\vert_{\psi}=
 -\derpar{H}{y^A}\Big\vert_{\psi}
 \label{hdw}
 \eeq
 which are known as the
 {\sl Hamilton-De Donder-Weyl equations}. But, as $H$ is a
 local Hamiltonian function, these equations are not covariant;
 that is, they transform in a non-trivial way under changes of
 coordinates (see \cite{CCI-91}).

 The way to overcome this problem (and get a system of covariant
 equations) consists in using a connection. In fact,
 a connection $\nabla$ in the bundle $\pi\colon E\to M$
 induces a linear section $j_\nabla\colon J^{1*}E\to{\cal M}\pi$
 of the projection $\mu$ \cite{CCI-91}, \cite{EMR-99b}.
 Then, we can construct the differentiable forms
 $$
 \Theta^\nabla:=j_\nabla^*\Theta \quad , \quad
 \Omega^\nabla:=-\d\Theta^\nabla=j_\nabla^*\Omega
 $$
 which are called the {\sl Liouville} $m$ and $(m+1)$ {\sl forms}
 of $J^{1*}E$ associated with the connection $\nabla$.
 Using natural coordinates in $J^{1*}E$ and ${\cal M}\pi$,
 if \dst\nabla = \d x^\mu\otimes\left(\derpar{}{x^\mu}+
 {\mit\Gamma}_\mu^A\derpar{}{y^A}\right)\),
 then we have that
 $j_\nabla(x^\mu,y^A,p^A_\mu)=
 (x^\mu,y^A,p^A_\mu,p=-p^A_\nu{\mit\Gamma}_A^\nu)$, and
 $$
 \Theta^\nabla =
 p_A^\mu\d y^A\wedge\d^{m-1}x_\mu-p_A^\mu{\mit\Gamma}^A_\mu\d^mx
 \quad , \quad
 \Omega^\nabla = -\d p_A^\mu\wedge\d y^A\wedge\d^{m-1}x_\mu+
 \d (p_A^\mu{\mit\Gamma}^A_\mu)\wedge\d^mx
 $$
 Now we have the following result:

 \begin{lem}
 If $h_1,h_2\colon J^{1*}E\to{\cal M}\pi$ are two sections of
 $\mu$, then $h_1^*\Theta-h_2^*\Theta=h_1-h_2$.
 \end{lem}
 \proof
 On the one hand,
 $h_1^*\Theta-h_2^*\Theta\in\df^m(J^{1*}E)$.
 On the other hand,
 $h_1-h_2\colon J^{1*}E\to{\cal M}\pi\equiv\Lambda^m_1\Tan^*E$
 has its image in $\pi^*\Lambda^m\Tan^*M$, because
 $h_1,h_2$ are sections of $\mu$. But we have a natural inclusion
 $\pi^*\Lambda^m\Tan^*M\hookrightarrow\Lambda^m\Tan^*J^{1*}E$
 given by means of the projection $\tau^1\colon J^{1*}E\to E$.
 Finally, the equality follows from a trivial calculation
 using natural coordinates.
 \qed

 Therefore, given a connection $\nabla$ and a
 Hamiltonian section $h$, from this Lemma we have that
 $$
 j_\nabla-h=j_\nabla^*\Theta-h^*\Theta\equiv\Theta^\nabla-\Theta_h:=
 {\cal H}^\nabla_h
 $$
 is a $\bar\tau^1$-semibasic $m$-form in $J^{1*}E$.
 It can be written as
 ${\cal H}^\nabla_h={\rm H}\bar\tau^{1^*}\omega$,
 where ${\rm H}\in\Cinfty (J^{1*}E)$ is the
 {\sl (global) Hamiltonian function} associated with
 ${\cal H}^\nabla_h$ and $\omega$.
 Then, we can define
 $$
 \Theta^\nabla_h:=\Theta^\nabla-{\cal H}^\nabla_h \quad ,\quad
 \Omega^\nabla_h:=-\d\Theta^\nabla_{\cal H}=
 \Omega^\nabla+\d{\cal H}^\nabla_h
 $$
 which are called the {\sl Hamilton-Cartan} $m$ and $(m+1)$ {\sl forms}
 of $J^{1*}E$ associated with the Hamiltonian section $h$
 and the connection $\nabla$.
 Their local expressions are
 \bea
 \Theta^\nabla_h &=& p_A^\mu\d y^A\wedge\d^{m-1}x_\mu -
 ({\rm H}+p^\mu_A{\mit\Gamma}^A_\mu)\d^mx \nonumber
 \\
 \Omega^\nabla_h &=& -\d p_A^\mu\wedge\d y^A\wedge\d^{m-1}x_\mu +
 \d ({\rm H} +p^\mu_A{\mit\Gamma}^A_\mu)\wedge\d^mx
 \label{omegaH}
 \eea
 where ${\rm H}$ is a global Hamiltonian function,
 whose relation with the local Hamiltonian function $H$
 associated with the Hamiltonian section $h$ is 
 ${\rm H}=H-p^A_\mu{\mit\Gamma}_A^\mu$ (in an open set $U$).
 In Field theory, every $\bar\tau^1$-semibasic $m$-form in $J^{1*}E$
 is usually called a {\sl Hamiltonian density}.

 As in the above case, the variational problem for the system
 $(J^{1*}E,\Omega^\nabla_h)$ leads to the following
 characterization of the critical sections
 \beq
 \psi^*\inn (X)\Omega^\nabla_h = 0 \quad ,\quad
 \mbox{for every} \ X\in\vf (J^{1*}E)
 \label{eqn0}
 \eeq
 which, in natural coordinates in $J^{1*}E$,
 is equivalent to the local equations (for the critical sections
 $\psi =(x^\mu,y^A(x),p^\mu_A(x))$)
 \beq
 \derpar{y^A}{x^\mu}\Big\vert_{\psi}=
 \left(\derpar{{\rm H}}{p^\mu_A}+{\mit\Gamma}^A_\mu\right)
 \Big\vert_{\psi} \quad ;\quad
 \derpar{p_A^\mu}{x^\mu}\Big\vert_{\psi}=
 - \left(\derpar{{\rm H}}{y^A}+
 p_B^\nu\derpar{{\mit\Gamma}^B_\nu}{y^A}\right)\Big\vert_{\psi}
 \label{he}
 \eeq
 which are covariant, and are called the {\sl Hamiltonian equations}
 of the system.

 If, conversely, we take a connection $\nabla$ and a
 Hamiltonian density ${\cal H}$, then making
 $j_\nabla-{\cal H}\equiv h_\nabla$ we obtain a section of $\mu$,
 that is, a Hamiltonian section,
 because ${\cal H}\colon J^{1*}E\to{\cal M}\pi$
 takes values in $\pi^*\Lambda^m\Tan^*M$.
 Hence we have proved the following:

 \begin{prop}
 A couple $(h,\nabla)$ in $J^{1*}E$ is equivalent to a couple
 $({\cal H},\nabla)$ (that is, given a connection $\nabla$,
 Hamiltonian sections and Hamiltonian densities are in
 one-to-one correspondence).
 \end{prop}

 Bearing in mind this last result, we have a third way of
 obtaining a Hamiltonian system, which consists in giving a couple
 $({\cal H},\nabla)$, and then define
 $$
 \Theta^\nabla_{\cal H}:=\Theta^\nabla-{\cal H} \quad ,\quad
 \Omega^\nabla_{\cal H}:=-\d\Theta^\nabla_{\cal H}=
 \Omega^\nabla+\d{\cal H}
 $$
 which are the {\sl Hamilton-Cartan} $m$ and $(m+1)$ {\sl forms}
 of $J^{1*}E$ associated with the Hamiltonian density ${\cal H}$
 and the connection $\nabla$.
 Their local expressions are the same as in (\ref{omegaH}),
 with ${\cal H}={\rm H}\bar\tau^{1*}\omega$.

 Summarizing, there are three ways of constructing Hamiltonian
 systems in Field theory, namely:
\begin{itemize}
  \item
  Giving a Hamiltonian section $h\colon J^{1*}E\to{\cal M}\pi$.
  \item
  Giving a couple $(h,\nabla)$, where $h$ is a Hamiltonian section
  and $\nabla$ a connection in $\pi\colon E\to M$.
  \item
  Giving a couple $({\cal H},\nabla)$, where ${\cal H}$
  is a Hamiltonian density.
\end{itemize}
 In each case, we can construct the Hamilton-Cartan forms
 and set a variational problem,
 which is called the {\sl Hamilton-Jacobi principle} of the
 Hamiltonian formalism. As we have said, the second and
 third way are equivalent.

 From now on, a couple $\hs$, or equivalently
 $(J^{1*}E,\Omega_{\cal H}^\nabla)$, will be called a
 {\sl Hamiltonian system}.

 \section{Hamiltonian equations, multivector fields and connections}

 We can set the Hamiltonian field
 equations using jet fields, connection forms and multivector fields
 (see the appendix \ref{mvf} for notation and terminology).

 First, an action of jet fields on forms
 is defined in the following way
 \cite{EMR-96}, \cite{EMR-98}:
 consider the bundle $J^1(J^{1*}E)$
 (the jet bundle of local sections of the projection $\bar\tau^1$),
 which is an affine bundle over
 $J^{1*}E$, whose associated vector bundle is
 $\bar\tau^{1*}\Tan^*M\otimes_E{\rm V}(\bar\tau^1 )$.
 We have
 $J^1(J^{1*}E) \stackrel{\tau^1_1}{\longrightarrow} J^{1*}E
  \stackrel{\bar\tau^1}{\longrightarrow} M$.
 If ${\cal Y}\colon J^{1*}E\to J^1(J^{1*}E)$
 is a jet field, a map $\bar{\cal Y}\colon\vf (M)\to\vf (J^{1*}E)$
 can be defined as follows: for every
 $Z\in\vf (M)$, $\bar{\cal Y}(Z)\in\vf(J^{1*}E)$ is the
 vector field given by
 $\bar{\cal Y}(Z)(\tilde y):=
 (\Tan_{\bar\tau^1(\tilde y)}\psi )(Z_{\bar\tau^1(\tilde y)})$,
 for every $\tilde y\in J^{1*}E$ and $\psi\in{\cal Y}(\tilde y)$.
 If ${\cal Y}\equiv
 (x^\mu,y^A,p_A^\mu,F_\mu^A(x,y,p),G_{A\mu}^\rho(x,y,p))$,
 its local expression is
 $$
 \bar{\cal Y} \left( f^\mu\derpar{}{x^\mu}\right)=
 f^\mu\left(\derpar{}{x^\mu}+F_\mu^A\derpar{}{y^A}
 +G_{A\mu}^\rho\derpar{}{p_A^\rho}\right)
 $$
 This map induces an action of ${\cal Y} $ on the forms in $J^{1*}E$.
 In fact, let $\xi\in\df^{m+k}(J ^{1*}E)$, with $k\geq 0$, we define
 $\inn ({\cal Y} )\xi\colon\vf (M)
 \times\stackrel{(m)}{\ldots}\times\vf (M)\longrightarrow\df^k(J^{1*}E)$
 given by
 $$
 [(\inn ({\cal Y} )\xi)(Z_1,\ldots ,Z_m)](\tilde y ;X_1,\ldots ,X_k):=
 \xi (\tilde y ;\bar{\cal Y}
 (Z_1),\ldots ,\bar{\cal Y}(Z_m),X_1,\ldots ,X_k)
 $$
 for $Z_1,\ldots ,Z_m\in\vf (M)$ and $X_1,\ldots ,X_k\in\vf (J^{1*}E)$.
 It is a $\Cinfty (M)$-linear and alternate map on the vector fields
 $Z_1,\ldots ,Z_m$. The $\Cinfty (J^{1*}E)$-linear map
 $\inn ({\cal Y} )$ so defined, extended by zero to forms of degree
 $p<m$, is called the {\sl inner contraction} with the jet field
 ${\cal Y}$. Then, it can be
 proved \cite{EMR-96}, \cite{EMR-98} that:

 \begin{lem}
 If ${\cal Y} $ is an integrable jet field and
 $\xi\in\df^{m+1}(J^{1*}E)$. Then $\inn ({\cal Y} )\xi =0$
 if, and only if, the integral sections $\psi\colon M\to J^{1*}E$
 of ${\cal Y}$ satisfy the relation $\psi^*\inn (X)\xi =0$,
 for every $X\in\vf (J^{1*}E)$.
 \end{lem}

 \begin{teor}
 Let $\hs$ be a Hamiltonian system.
 The critical sections of the Hamilton-Jacobi principle are
 the sections $\psi\in\Gamma_c(M,J^{1*}E)$ satisfying any one of
 the following conditions:
 \ben
 \item
  They are the integral sections of an integrable jet field
 ${\cal Y}_{\cal H}\colon J^{1*}E\to J^1(J^{1*}E)$ satisfying that
 $\inn({\cal Y}_{\cal H})\Omega^\nabla_h=0$.
 \item
 They are the integral sections of an integrable connection
 $\nabla_{\cal H}$ satisfying that
 $\inn (\nabla_{\cal H})\Omega^\nabla_h=
 (m-1)\Omega^\nabla_h$.
 \item
 They are the integral sections of a class of integrable and
 $\bar\tau^1$-transverse multivector fields
 $\{ X_{\cal H}\}\subset\vf^m(J^{1*}E)$ such that
 $\inn (X_{\cal H})\Omega^\nabla_h=0$,
 for every $X_{\cal H}\in\{ X_{\cal H}\}$.
 \een
 \label{hameq}
 \end{teor}
 \proof
 Critical sections are characterized by the equation (\ref{eqn0}).
 Then, using the above lemma with
 $\xi\equiv\Omega_h^\nabla$,
 we obtain the equivalence between (\ref{eqn0}) and the item 1.

 For the second item it suffices to use the expression in natural
 coordinates of a connection
 $$ \nabla_{\cal H}=
 \d x^\mu\otimes \left(\derpar{}{x^\mu}+F_\mu^A\derpar{}{y^A}+
 G^\rho_{A\mu}\derpar{}{p^\rho_A}\right) $$
 Hence, bearing in mind
 the local expression (\ref{omegaH}), we prove that the condition
 $\inn (\nabla_{\cal H})\Omega^\nabla_h=(m-1)\Omega^\nabla_h$ holds
 for an integrable connection if, and only if, the Hamiltonian
 equations (\ref{he}) hold for its integral sections (see
 \cite{LMM-95} and \cite{Sa-89}).

 Finally, item 3 is a direct consequence of the
 equivalence between orientable and integrable jet fields
 ${\cal Y}\colon J^{1*}E\to J^1(J^{1*}E)$, and classes of locally
 decomposable, $\bar\tau^1$-transverse and integrable multivector
 fields $\{ X\}\subset\vf^m(J^{1*}E)$.
 \qed

Thus, in Hamiltonian Field theories we search for (classes of)
$\bar\tau^1$-transverse and locally decomposable multivector
fields $X_{\cal H}\in\vf^m(J^{1*}E)$ such that:
 \ben
 \item
 The equation $\inn (X_{\cal H})\Omega^\nabla_h=0$ holds.
 \item
 $X_{\cal H}$ are integrable.
 \een
 A representative of the class of multivector fields satisfying
 the first condition can be selected by demanding that
 $\inn (X_{\cal H})(\bar\tau^{1*}\omega)=1$.
 Then its local expression is
 \beq
 X_{\cal H}=\bigwedge_{\mu=1}^m
 \left(\derpar{}{x^\mu}+F_\mu^A\derpar{}{y^A}+
 G^\rho_{A\mu}\derpar{}{p^\rho_A}\right)
 \label{locmvf}
 \eeq
 Concerning to the second condition,
 let us recall that, if $\{ X_{\cal H}\}\subset\vf^m(J^{1*}E)$ is a
 class of locally decomposable and $\bar\tau^1$-transverse
 multivector fields, then $X_{\cal H}$ is integrable if, and only if,
 the curvature of the connection associated
 with this class vanishes everywhere.

 \begin{definition}
 $X_{\cal H}\in\vf^m(J^{1*}E)$ will be called a
 {\rm Hamilton-De Donder-Weyl (HDW) multivector field}
 for the system $\hs$ if it is
 $\bar\tau^1$-transverse, locally decomposable and verifies the
 equation $\inn (X_{\cal H})\Omega^\nabla_h=0$.

 We denote the set of HDW-multivector fields by $\vf^m_{HDW}\hs$.
 \end{definition}

 \begin{teor}
 {\rm (Existence and local multiplicity of HDW-multivector
 fields):}
 Let $\hs$ be a Hamiltonian system.
 \ben
 \item
 There exist classes of HDW-multivector
 fields $\{ X_{\cal H}\}\subset\vf^m_{HDW}(J^{1*}E)$,
(and hence equivalent jet fields
 ${\cal Y}_{\cal H}\colon J^1E\to J^1(J^{1*}E)$ with associated connection
 forms $\nabla_{\cal H}$, satisfying that
 $\inn ({\cal Y}_{\cal H})\Omega^\nabla_h=0$
 and $\inn (\nabla_{\cal H})\Omega^\nabla_h=
 (m-1)\Omega^\nabla_h$, respectively).
 \item
 In a local system the above solutions depend on $N(m^2-1)$
 arbitrary functions. \een \label{holsecreg}
 \end{teor}
 \proof
 \ben
 \item
 First we analyze the local existence of solutions and then their
 global extension.

In a chart of natural coordinates in $J^{1*}E$, the expression of
$\Omega^\nabla_h$ is (\ref{omegaH}); and taking the multivector
field given in (\ref{locmvf}) as representative of the class $\{
X_{\cal H}\}$, from the relation $\inn (X_{\cal
H})\Omega^\nabla_h=0$ we obtain the following conditions: \bit
 \item
 The coefficients on $\d p_A^\mu$ must vanish:
 \beq
 0=F^A_\nu -\derpar{{\rm H}}{p_A^\nu}-{\mit\Gamma}^A_\nu
 \qquad (\mbox{for every $A,\nu$})
 \label{eqsG1}
 \eeq
 This system of $Nm$ linear equations determines
 univocally the functions $F^A_\nu$.
 \item
 The coefficients on $\d y^A$ must vanish
 \beq
 0=G^\mu_{A\mu}+\derpar{{\rm H}}{y^A}+
 p_B^\nu\derpar{{\mit\Gamma}^B_\nu}{y^A}
 \qquad (A=1,\ldots ,N)
 \label{eqsG2}
 \eeq
 which is a compatible system of $N$ linear equations on the
 $Nm^2$ functions $G^\mu_{A\nu}$.
 \item
 Using these results we obtain that the coefficients on
 $\d x^\mu$ vanish identically.
 \eit
 These results allow us to assure
 the local existence of (classes of) multivector fields satisfying
 the desired conditions. The corresponding global solutions are
 then obtained using a partition of unity subordinated to a
 cover of $J^{1*}E$ made of natural charts.

 (Note that, if
 $\psi =(x^\mu ,y^A(x^\nu ),p^\mu_A(x^\nu ))$
 is an integral section of $X_{\cal H}$
 (resp. ${\cal Y}_{\cal H}$), then
 $$
 F^A_\mu\circ \psi =\derpar{y^A}{x_\mu} \quad ; \quad
 G^\mu_{A\mu}\circ \psi = -\derpar{p^\mu_A}{x^\mu}
 $$
 and then equations (\ref{eqsG1}) and
 (\ref{eqsG2}) are the Hamiltonian equations for $\psi$).
 \item
 In natural coordinates in $J^{1*}E$,
 a HDW-multivector field $X_{\cal H}\in\{ X_{\cal H}\}$ is
 given by (\ref{locmvf}). So, it is determined by the $Nm$
 coefficients $F^A_\nu$ (which are obtained as the solution of
 (\ref{eqsG1})), and by the $Nm^2$ coefficients
 $G^\mu_{A\nu}$, which are related  by the $N$ independent
 equations (\ref{eqsG2}). Therefore, there are $N(m^2-1)$ arbitrary
 functions. \een \qed

 Finally we try to determine if it is possible
 to find a class of integrable
 HDW-multivector fields. Hence we must impose
 that the corresponding multivector field $X_{\cal H}$ verify the
 integrability condition; that is, the curvature of the associated
 connection $\nabla_{\cal H}$ vanishes everywhere, that is,
 \beann
 0 &=&
 \left(\derpar{F_\eta^B}{x^\mu}+F_\mu^A\derpar{F_\eta^B}{y^A}+
 G^\gamma_{B\mu}\derpar{F_\eta^B}{p^\gamma_A}-
 \derpar{F_\mu^B}{x^\eta}-F_\eta^A\derpar{F_\mu^B}{y^A}-
 G^\rho_{A\eta}\derpar{F_\mu^B}{p^\rho_A}\right)
 (\d x^\mu\wedge\d x^\eta )\otimes\derpar{}{y^B}+
 \\ & &
 \left(\derpar{G^\rho_{B\eta}}{x^\mu}+F_\mu^A\derpar{G^\rho_{B\eta}}{y^A}+
 G^\gamma_{A\mu}\derpar{G^\rho_{B\eta}}{p^\gamma_A}-
 \derpar{G^\rho_{B\mu}}{x^\eta}-F_\eta^A\derpar{G^\rho_{B\mu}}{y^A}-
 G^\gamma_{A\eta}\derpar{G^\rho_{B\mu}}{p^\gamma_A}\right)
 (\d x^\mu\wedge\d x^\eta )\otimes\derpar{}{p^\rho_B}
 \eeann
 or, what is equivalent, the
 following system of equations hold (for $1\leq\mu <\eta\leq m$)
 \bea 0 &=&
 \derpar{F_\eta^B}{x^\mu}+F_\mu^A\derpar{F_\eta^B}{y^A}+
 G^\gamma_{A\mu}\derpar{F_ \eta^B}{p^\gamma_A}-
 \derpar{F_\mu^B}{x^\eta}-F_\eta^A\derpar{F_\mu^B}{y^A}-
 G^\rho_{A\eta}\derpar{F_\mu^B}{p^\rho_A} \nonumber
 \\ & =&
 \frac{\partial^2\tilde{\rm H}}{\partial x^\mu\partial p_B^\eta}+
 \derpar{H}{p_A^\mu}
 \frac{\partial^2 H}{\partial y^A\partial p_B^\eta}+
 \nonumber \\ & &
 G^\gamma_{A\mu}\frac{\partial^2 H}{\partial
 p_A^\gamma\partial p_B^\eta}-
 \frac{\partial^2 H}{\partial x^\eta\partial p_B^\eta}-
 \derpar{H}{p_A^\eta}\frac{\partial^2 H}{\partial y^A\partial
 p_B^\mu}- G^\rho_{A\eta}\frac{\partial^2 H}{\partial
 p_A^\rho\partial p_B^\mu} \label{curvcero1}
 \\
 0 &=&
 \derpar{G^\rho_{B\eta}}{x^\mu}+F_\mu^A\derpar{G^\rho_{B\eta}}{y^A}+
 G^\gamma_{A\mu}\derpar{G^\rho_{B\eta}}{p^\gamma_A}-
 \derpar{G^\rho_{B\mu}}{x^\eta}-F_\eta^A\derpar{G^\rho_{B\mu}}{y^A}-
 G^\gamma_{A\eta}\derpar{G^\rho_{B\mu}}{p^\gamma_A} \nonumber
 \\ &=&
 \derpar{G^\rho_{B\eta}}{x^\mu}+
 \derpar{H}{p^\mu_A}\derpar{G^\rho_{B\eta}}{y^A}+
 G^\gamma_{A\mu}\derpar{G^\rho_{B\eta}}{p^\gamma_A}-
 \derpar{G^\rho_{B\mu}}{x^\eta}-
 \derpar{H}{p^\eta_A}\derpar{G^\rho_{B\mu}}{y^A}-
 G^\gamma_{A\eta}\derpar{G^\rho_{B\mu}}{p^\gamma_A}
 \label{curvcero2}
 \eea
 (where $H\equiv{\rm H}+p^A_\mu{\mit\Gamma}_A^\mu$,
 and use is made of the Hamiltonian equations).
 Since these additional conditions on the
 functions $G^\mu_{A\nu}$ must be imposed in order to assure that
 $X_{\cal H}$ is integrable, the number of arbitrary functions will
 be in general less than $N(m^2-1)$.

 As far as we know, since this is a system of partial differential
 equations with linear restrictions, there is no way of assuring
 the existence of an integrable solution, or of selecting it.
 Observe that, considering the Hamiltonian equations for the
 coefficients $G^\mu_{A\nu}$(equations (\ref{eqsG2})), together
 with the integrability conditions (\ref{curvcero1}) and
 (\ref{curvcero2}), we have \dst N+\frac{1}{2}Nm(m-1)\) linear
 equations and \dst\frac{1}{2}Nm^2(m-1)\) partial differential
 equations. Then, if the set of linear restrictions (\ref{eqsG2})
 and (\ref{curvcero1}) allow us to isolate \dst
 N+\frac{1}{2}Nm(m-1)\) coefficients $G^\mu_{A\nu}$ as functions on
 the remaining ones; and the set of \dst\frac{1}{2}Nm^2(m-1)\)
 partial differential equations (\ref{curvcero2}) on these
 remaining coefficients satisfies the conditions on {\sl
 Cauchy-Kowalewska's theorem} \cite{Di-74}, then the existence of
 integrable HDW-multivector fields (in $J^{1*}E)$) is assured.
 If this is not the case, we can eventually select some particular
 HDW-multivector field solution, and apply the integrability
 algorithm developed in \cite{EMR-98} in order to find a
 submanifold ${\cal I}\hookrightarrow J^{1*}E$ (if it exists),
 where this multivector field is integrable
 (and tangent to ${\cal I}$).

 Other results concerning the expression of the Hamiltonian
 equations in terms of multivector fields can be found in
 \cite{Ka-93}, \cite{Ka-95}, \cite{Ka-97b} and \cite{Ka-98}, where
 the definition of Poisson algebras in Field theories is also given
 (see also \cite{CCI-91}).

 \section{Symmetries and first integrals}
 \protect\label{ntmvf}

 Next we recover the idea of {\sl first integral}
 or {\sl conserved quantity}, and state
 Noether's theorem for Hamiltonian systems in Field theory,
 in terms of multivector fields.
 In this sense, a great part of our discussion is a generalization of
 the results obtained for non-autonomous (non-regular) mechanical systems
 (see, in particular, \cite{LM-96}, and references quoted therein).
 We refer to appendix \ref{mvf}
 to review the definition of the basic differential
 operations on the set of multivector fields in a manifold.

 Consider a Hamiltonian system $\hs$. Let
 $$
 \ker^m\,\Omega^\nabla_h:=\{ {\cal Z}\in\vf^m(J^{1*}E)
 \ ; \ \inn({\cal Z})\Omega^\nabla_h=0\}
 $$
 and let $\ker^m_\omega\,\Omega^\nabla_h\subset\vf^m(J^{1*}E)$
 be the set of $m$-multivector fields satisfying that
 \beq
 \inn(X)\Omega^\nabla_h=0 \quad ,\quad
 \inn(X)(\bar\tau^{1*}\omega)\not=0
 \label{generic}
 \eeq
 These are $\bar\tau^1$-transverse multivector fields (but not
 locally decomposable, necessarily), and as usual
 we can select a representative on each
 equivalence class of solutions, by demanding that
 $\inn(X)(\bar\tau^{1*}\omega)=1$.
 Remember that HDW-multivector fields are
 solutions of (\ref{generic}) which are locally decomposable.
 Then, if $\vf^m_{IHDW}\hs$
 denotes the set of integrable HDW-multivector fields,
 we obviously have that
 $$
 \vf^m_{IHDW}\hs\subset
 \vf^m_{HDW}\hs\subset\ker^m_\omega\,\Omega^\nabla_h
 \subset\ker^m\,\Omega^\nabla_h
 $$
 Now we introduce the following terminology \cite{Gc-73}, \cite{LM-96}:

 \begin{definition}
 A {\rm first integral} or a {\rm conserved quantity}
 of a Hamiltonian system $\hs$
 is a form $\xi\in\df^{m-1}(J^{1*}E)$ such that
 $\Lie(X)\xi=0$, for every
 $X\in\ker^m_\omega\,\Omega^\nabla_h$.
 \end{definition}

 Observe that, in this case,
 $\Lie(X)\xi=(-1)^{m+1}\inn(X)\d\xi$.

 \begin{prop}
 If $\xi\in\df^{m-1}(J^{1*}E)$ is a
 first integral of a Hamiltonian system $\hs$, and
 $X\in\ker^m_\omega\,\Omega^\nabla_h$ is integrable,
 then $\xi$ is closed on the integral submanifolds of $X$.
 That is, if $j_S\colon S\hookrightarrow J^{1*}E$
 is an integral submanifold of $X$, then $\d j_S^*\xi=0$.
 \end{prop}
 \proof
 Let $X_1,\ldots ,X_m\in\vf (J^{1*}E)$ be independent vector fields
 tangent to the ($m$-dimensional) integral submanifold $S$. Then
 $X=fX_1\wedge\ldots\wedge X_m$, for some $f\in\Cinfty (J^{1*}E)$.
 Therefore, as $\inn(X)\d\xi=0$, we have that
 $$
 j_S^*[\d\xi (X_1,\ldots ,X_m)]=
 j_S^*\inn(X_1\wedge\ldots\wedge X_m)\d\xi=0
 $$
 \qed

 Conserved quantities can be characterized as follows:

 \begin{prop}
 If $\xi\in\df^{m-1}(J^{1*}E)$
 is a first integral of a Hamiltonian system $\hs$, then
 $\Lie({\cal Z})\xi=0$, for every
 ${\cal Z}\in\ker^m\,\Omega^\nabla_h$.
 \label{fintchar}
 \end{prop}
 \proof
 Consider the conditions (\ref{generic}), with
 $\inn(X)(\bar\tau^{1*}\omega)=1$, and
 let $X_0\in\ker^m_\omega\,\Omega^\nabla_h$
 be a particular solution.
 Then, any other solution can be obtained by making
 $fX_0+Z$, with $Z\in\ker^m\,\Omega^\nabla_h\cap
 \ker^m\, (\bar\tau^{1*}\omega)$ and $f\in\Cinfty (J^{1*}E)$.
 Thus we have that
 $$
 \ker^m_\omega\,\Omega^\nabla_h =
 \{fX_0+\ker^m\,\Omega^\nabla_h\cap
 \ker^m\, (\bar\tau^{1*}\omega) \ ; \
 f\in\Cinfty(J^{1*}E)\} \subset\ker\,\Omega^\nabla_h
 $$
 Then, for every $Z\in\ker^m\,\Omega^\nabla_h\cap
 \ker^m\, (\bar\tau^{1*}\omega)$, we have that
 $Z=X_1-X_2$, with $X_1,X_2\in\ker^m_\omega\,\Omega^\nabla_h$
 such that
 $\inn(X_1)(\bar\tau^{1*}\omega)=\inn(X_2)(\bar\tau^{1*}\omega)$.
 Hence, if $\xi$ is a first integral, we have that
 $\Lie(Z)\xi=0$.
 On the other hand, taking
 $X_0\in\ker^m_\omega\,\Omega^\nabla_h$,
 for every ${\cal Z}\in\ker^m\,\Omega^\nabla_h$
 we can write the identity
 $$
 {\cal Z}=({\cal Z}-\inn({\cal Z})(\bar\tau^{1*}\omega)X_0)+
 \inn({\cal Z})(\bar\tau^{1*}\omega)X_0
 $$
 then, if $\inn(X_0)(\bar\tau^{1*}\omega)=1$, it follows that
 ${\cal Z}-\inn({\cal Z})(\bar\tau^{1*}\omega)X_0\in
 \ker^m\,\Omega^\nabla_h\cap
 \ker^m\, (\bar\tau^{1*}\omega)$, hence
 $$
 \Lie({\cal Z})\xi=
 \Lie({\cal Z}-\inn({\cal Z})(\bar\tau^{1*}\omega)X_0)\xi+
 \Lie(\inn({\cal Z})(\bar\tau^{1*}\omega)X_0)\xi=
 (-1)^{m+1}\inn({\cal Z})(\bar\tau^{1*}\omega)\inn(X_0)\d\xi=0
 $$
 since $\d\inn(X_{\cal H})\xi_Y=0$, because
 $\xi_Y\in\df^{m-1}(J^{1*}E)$.
 \qed

 The converse of this statement holds obviously, and
 hence this is a characterization of first integrals.

 Next we introduce the following terminology
 (which will be justified in Theorem \ref{gsymsol}):

 \begin{definition}
 An {\rm (infinitesimal) general symmetry}
 of a Hamiltonian system $\hs$ is a vector field $Y\in\vf (J^{1*}E)$
 satisfying that
 $[Y,\ker^m\,\Omega^\nabla_h]\subset\ker^m\,\Omega^\nabla_h$.
 \end{definition}

 Bearing in mind the properties of multivector fields (see the Appendix),
 we obtain that general symmetries have the following basic properties:
 \begin{itemize}
  \item
 If $Y\in\vf (J^{1*}E)$ is a general symmetry,
 then so is $Y+Z$, for every $Z\in\ker\,\Omega^\nabla_h$.
  \item
 If $Y_1,Y_2\in\vf (J^{1*}E)$ are general symmetries,
 then so is $[Y_1,Y_2]$.
 \end{itemize}

 A first characterization of general symmetries is given by:

 \begin{lem}
 Let $\hs$ be a Hamiltonian system, $Y\in\vf(J^{1*}E)$,
 and let $F_t$ be a local flow of $Y$.
 $Y$ is a general symmetry if, and only if,
 $F_{t*}(\ker^m\,\Omega^\nabla_h)\subset
 \ker^m\,\Omega^\nabla_h$,
 in the corresponding open sets.
 \label{previo0}
 \end{lem}
 \proof
 As $\ker^m\,\Omega^\nabla_h$
 is locally finite-generated, we can take a local basis
 $\moment{Z}{1}{r}$ of $\ker^m\,\Omega^\nabla_h$,
 and then the assertion is equivalent to proving that
 $[Y,Z_i]=f_i^jZ_j$ if, and only if, $F_{t*}Z_i=g_i^jZ_j$
 (for every $i=1,\ldots ,r$), where $g_i^j$ are differentiable
 functions on the corresponding open set,
 also depending on $t$.

 It is clear that, if $F_{t*}Z_i=g_i^jZ_j$, then $[Y,Z_i]=f_i^jZ_j$.

 For the converse, we have to prove the existence of functions
 $g_i^j$ such that $F_{t*}Z_i=g_i^jZ_j$.
 Suppose that $[Y,Z_i]=f_i^jZ_j$, and remember that
 \dst\frac{\d}{\d t}\Big\vert_{t=s}F_{t*}Z_i=F_{s*}[Y,Z_i]\) .
 Hence, on the one hand we obtain
 $$
 F_{s*}[Y,Z_i] =
 F_{s*}(f_i^jZ_j)=(F_s^{-1})^*f_i^jF_{s*}Z_j=
 (F_s^{-1})^*f_i^j(g_j^kZ_k)
 $$
 and on the other hand, we have that
 $$
 \frac{\d}{\d t}\Big\vert_{t=s}F_{t*}Z_i=
 \frac{\d}{\d t}\Big\vert_{t=s}g_i^kZ_k=
 \frac{\d g_i^k}{\d t}\Big\vert_{t=s}Z_k
 $$
 therefore, comparing these expressions, we conclude that
 $$
 \frac{\d g_i^k}{\d t}=(F_t^{-1})^*f_i^jg_j^k
 $$
 This is a system of ordinary linear differential equations
 for the functions $g_i^k$,
 which, with the initial condition $g_i^k(0)=\delta_i^k$,
 has a unique solution, defined for every $t$ on the domain of
 $F_t$. Then, taking this solution, the result holds.
 \qed

 Using this Lemma, we can prove that:

 \begin{teor}
 Let $Y\in\vf (J^{1*}E)$ be a
 general symmetry of a Hamiltonian system $\hs$, and
 $F_t$ a local flow of $Y$.
 \ben
 \item
 If ${\cal Z}\in\ker^m\,\Omega^\nabla_h$ is an integrable
 multivector field, then $F_t$ transforms integral submanifolds
 of ${\cal Z}$ into integral submanifolds of $F_{t*}{\cal Z}$.
 \item
 In particular,
 if $Y\in\vf (J^{1*}E)$ is $\bar\tau^1$-projectable, and
 $X_{\cal H}\in\vf^m_{IHDW}\hs$,
 then $F_t$ transforms critical sections of $X_{\cal H}$
 into critical sections of $F_{t*}X_{\cal H}$, and hence
 $F_{t*}X_{\cal H}\in\vf^m_{IHDW}\hs$.
 \een
 \label{gsymsol}
 \end{teor}
 \proof
 \ben
 \item
 Let $X_1,\ldots ,X_m\in\vf (J^{1*}E)$ be vector fields
 locally expanding the involutive distribution associated with
 ${\cal Z}$. Then $F_{t*}X_1,\ldots ,F_{t*}X_m$ generate another
 distribution which is also involutive, and, hence, is associated
 with a class of locally decomposable multivector fields
 whose representative is just $F_{t*}{\cal Z}$, by construction.
 The assertion about the integral submanifolds is then immediate.
 \item
 First observe that, as $Y$ is $\bar\tau^1$-projectable,
 then $F_t$ restricts to a local flow $F_t^M$ in $M$; that is,
 we have $F_t^M\circ\bar\tau^1=\bar\tau^1\circ F_t$.
 Now, for every $\psi\colon M\to J^{1*}E$, integral section of
 $X_{\cal H}$, we can define $\psi_t\colon M\to J^{1*}E$
 by the relation $F_t\circ\psi=\psi_t\circ F_t^M$,
 which is also a section of $\bar\tau^1$, because
 $$
 \bar\tau^1\circ\psi_t =
 \bar\tau^1\circ F_t\circ\psi\circ (F_t^M)^{-1}=
 F_t^M\circ\bar\tau^1\circ\psi\circ (F_t^M)^{-1}=
 F_t^M\circ (F_t^M)^{-1}= {\rm Id}_M
 $$
 since $\bar\tau^1\circ\psi={\rm Id}_M$.
 Then, observe that, by construction,
 ${\rm Im}\,\psi_t=F_t({\rm Im}\,\psi)$ is
 an integral submanifold of $F_{t*}X_{\cal H}$,
 and as is a section of $\bar\tau^1$, it is
 $\bar\tau^1$-transverse. Hence $F_{t*}X_{\cal H}$
 (which belongs to $\ker^m\,\Omega^\nabla_h$,
 by Lemma \ref{previo0}) is integrable (then locally decomposable),
 and as its integral submanifolds are sections
 of $\bar\tau^1$, it is $\bar\tau^1$-transverse,
 thus $F_{t*}X_{\cal H}\in\vf^m_{IHDW}\hs$.
 \een
 \qed

 General symmetries can be used for obtaining
 conserved quantities, as follows:

 \begin{prop}
 If $\xi\in\df^{m-1}(J^{1*}E)$ is a first integral
 of a Hamiltonian system $\hs$, then
 so is $\Lie(Y)\xi$, for every general symmetry
 $Y\in\vf (J^{1*}E)$.
 \end{prop}
 \proof
 For every first integral $\xi\in\df^{m-1}(J^{1*}E)$,
 and ${\cal Z}\in\ker^m\,\Omega^\nabla_h$,
 if $Y\in\vf (J^{1*}E)$ is a general symmetry, we have that
 $$
 \Lie({\cal Z})\Lie(Y)\xi =
 \Lie([{\cal Z},Y])\xi+\Lie(Y)\Lie({\cal Z})\xi=
 \Lie([{\cal Z},Y])\xi=0
 $$
 since $[{\cal Z},Y]\in\ker^m\,\Omega^\nabla_h$,
 and as a consequence of Proposition \ref{fintchar}.
 \qed

 \section{Noether's theorem for multivector fields}
 \protect\label{ntmvfbis}

 There is another kind of symmetries which play a relevant role,
 as generators of conserved quantities:

 \begin{definition}
 An {\rm (infinitesimal) Cartan} or {\rm Noether symmetry}
 of a Hamiltonian system $\hs$ is a vector field $Y\in\vf (J^{1*}E)$
 satisfying that $\Lie(Y)\Omega^\nabla_h=0$.
 \label{CNsym}
 \end{definition}

 {\bf Remarks}:
 \bit
 \item
 It is immediate to prove that, if $Y_1,Y_2\in\vf (J^{1*}E)$
 are Cartan-Noether symmetries, then so is $[Y_1,Y_2]$.
 \item
 Observe that the condition $\Lie(Y)\Omega^\nabla_h=0$
 is equivalent to demanding that $\inn(Y)\Omega^\nabla_h$
 is a closed $m$-form in $J^{1*}E$. Therefore, for every
 $p\in J^{1*}E$, there exists an open neighborhood $U_p\ni p$,
 and $\xi_Y\in\df^{m-1}(U_p)$, such that
 $\inn(Y)\Omega^\nabla_h=\d\xi_Y$ (on $U_p$).
 Thus, a Cartan-Noether symmetry of a Hamiltonian system is
 just a {\sl locally Hamiltonian vector field}
 for the multisymplectic form $\Omega^\nabla_h$,
 and $\xi_Y$ is the corresponding {\sl local Hamiltonian form},
 which is unique, up to a closed $(m-1)$-form.
 \eit

 Cartan-Noether symmetries have the following property:

 \begin{prop}
 Let $Y\in\vf (J^{1*}E)$ be a Cartan-Noether symmetry
 of a Hamiltonian system $\hs$. Therefore:
 \ben
 \item
 $\Lie(Y)\Theta^\nabla_h$ is a closed form,
 hence, in an open set $U\subset J^{1*}E$, there exist
 $\zeta_Y\in\df^{m-1}(U)$ such that $\Lie(Y)\Theta^\nabla_h=\d\zeta_Y$.
 \item
 If $\inn(Y)\Omega^\nabla_h=\d\xi_Y$,
 in an open set $U\subset J^{1*}E$, then
 $$
 \Lie(Y)\Theta^\nabla_h=\d (\inn(Y)\Theta^\nabla_h-\xi_Y)=\d\zeta_Y
 \quad \mbox{\rm (in $U$)}
 $$
 \een
 \label{xizeta}
 \end{prop}
 \proof
 \ben
 \item
 The first item is immediate since
 $\d\Lie(Y)\Theta^\nabla_h=\Lie(Y)\d\Theta^\nabla_h=0$.
 \item
 For the second item we have
 $$
 \Lie(Y)\Theta^\nabla_h=
 \d\inn(Y)\Theta^\nabla_h+\inn(Y)\d\Theta^\nabla_h=
 \d\inn(Y)\Theta^\nabla_h-\inn(Y)\Omega^\nabla_h=
 \d (\inn(Y)\Theta^\nabla_h-\xi_Y)
 $$
 Hence we can write $\xi_Y=\inn(Y)\Theta^\nabla_h-\zeta_Y$
 (up to a closed $(m-1)$-form).
 \een
 \qed

 {\bf Remark}:
 \bit
 \item
 As a particular case, if for a Cartan-Noether symmetry $Y$
 the condition $\Lie (Y)\Theta^\nabla_h=0$ holds,
 we can take
 $\xi_Y=\inn(Y)\Theta^\nabla_h$.
 In this case $Y$ is said to be an {\sl exact Cartan-Noether symmetry}.
 \eit

 Cartan-Noether symmetries and general symmetries are closely
 related. In fact:

 \begin{prop}
 Every Cartan-Noether symmetry of a Hamiltonian system $\hs$
 is a general symmetry.
 \end{prop}
 \proof
 Let $Y\in\vf(J^{1*}E)$ be a Cartan-Noether symmetry.
 For every ${\cal Z}\in\ker^m\,\Omega^\nabla_h$, we have that
 $$
 \inn([Y,{\cal Z}])\Omega^\nabla_h=
 \Lie(Y)\inn({\cal Z})\Omega^\nabla_h+
 (-1)^{2+m}\inn({\cal Z})\Lie(Y)\Omega^\nabla_h=0
 $$
 and therefore
 $[Y,{\cal Z}]\subset\ker^m\,\Omega^\nabla_h$.
 \qed

 Finally, the classical {\sl Noether's theorem}
 of Hamiltonian mechanics can be generalized to Field theory as follows:

 \begin{teor}
 {\rm (Noether):}
 If $Y\in\vf (J^{1*}E)$ is a Cartan-Noether symmetry of a Hamiltonian
 system $\hs$, with $\inn(Y)\Omega^\nabla_h=\d\xi_Y$. Then,
 for every HDW-multivector field
 $X_{\cal H}\in\vf^m(J^{1*}E)$, we have that
 $$
 \Lie(X_{\cal H})\xi_Y=0
 $$
 that is, any Hamiltonian $(m-1)$-form
 $\xi_Y$ associated with $Y$ is a first integral of $\hs$.
 \label{Nth}
 \end{teor}
 \proof
 If $Y\in\vf (J^{1*}E)$ is a Cartan-Noether symmetry then
 $$
 \Lie(X_{\cal H})\xi_Y=\d\inn(X_{\cal H})\xi_Y
 -(-1)^m\inn(X_{\cal H})\d\xi_Y=
 -(-1)^m\inn(X_{\cal H})\inn(Y)\Omega^\nabla_h=
 -\inn(Y)\inn(X_{\cal H})\Omega^\nabla_h=0
 $$
 \qed

 It is interesting to remark that, to our knowledge,
 given a first integral of a Hamiltonian system,
 there is no a straightforward way of associating to it
 a Cartan-Noether symmetry $Y$. The main
 obstruction is that, given a $(m-1)$-form $\xi$,
 the existence of a solution for the equation
 $\inn(Y)\Omega^\nabla_h=\d\xi$ is not assured
 (even in the case $\Omega^\nabla_h$ being 1-nondegenerate).
 Hence, in general, the {\sl converse Noether theorem} cannot be stated
 for multisymplectic Hamiltonian systems.

 Noether's theorem associates first integrals to Cartan-Noether
 symmetries. But these kinds of symmetries do not exhaust the set of
 (general) symmetries. As is known, in mechanics there are
 dynamical symmetries which are not of Cartan type,
 which generate also conserved quantities
 (see \cite{LMR-99}, \cite{Ra-95}, \cite{Ra-97},
 for some examples). These are the so-called {\sl hidden
 symmetries}. Different attempts have been made
 to extend Noether's theorem in order
 to include these symmetries and the corresponding conserved
 quantities. Next we present a generalization of the Noether
 theorem \ref{Nth}, which is based in the approach of
 reference \cite{SC-81} for mechanical systems.

 First we introduce the {\sl higher-order Cartan-Noether symmetries},
 generalizing the definition \ref{CNsym} in the following way:

 \begin{definition}
 An {\rm (infinitesimal) Cartan-Noether symmetry of order $n$}
 of a Hamiltonian system $\hs$ is a vector field $Y\in\vf (J^{1*}E)$
 satisfying that:
 \ben
 \item
 $Y$ is a general symmetry.
 \item
 $\Lie^n(Y)\Omega^\nabla_h=0$, but
 $\Lie^k(Y)\Omega^\nabla_h\not= 0$, for $k<n$.
 \een
 \label{CNsymn}
 \end{definition}

 Observe that Cartan-Noether symmetries of order $n>1$ are not
 necessarily Hamiltonian vector fields for the multisymplectic form
 $\Omega^\nabla_h$. Nevertheless we have that:

 \begin{prop}
 If $Y\in\vf (J^{1*}E)$ is a Cartan-Noether symmetry of order $n$
 of a Hamiltonian system $\hs$, then the form
 $\Lie^{n-1}(Y)\inn(Y)\Omega^\nabla_h\in\df^m(J^{1*}E)$
 is closed.
 \end{prop}
 \proof
 In fact, from the definition \ref{CNsymn} we obtain
 $$
 0=\Lie^n(Y)\Omega^\nabla_h=
 \Lie^{n-1}(Y)\Lie(Y)\Omega^\nabla_h=
 \Lie^{n-1}(Y)\d\inn(Y)\Omega^\nabla_h=
 \d\Lie^{n-1}(Y)\inn(Y)\Omega^\nabla_h
 $$
 \qed

 Hence, this condition is equivalent to demanding that, for every
 $p\in J^{1*}E$, there exists an open neighborhood $U_p\ni p$,
 and $\xi_Y\in\df^{m-1}(U_p)$, such that
 $\Lie^{n-1}(Y)\inn(Y)\Omega^\nabla_h=\d\xi_Y$ (on $U_p$).
 Then, the result stated in Proposition \ref{xizeta}
 can be generalized as follows:

 \begin{prop}
 Let $Y\in\vf (J^{1*}E)$ be a Cartan-Noether symmetry of order $n$
 of a Hamiltonian system $\hs$. Therefore:
 \ben
 \item
 $\Lie^n(Y)\Theta^\nabla_h$ is a closed form,
 hence, in an open set $U\subset J^{1*}E$, there exist
 $\zeta_Y\in\df^{m-1}(U)$ such that  $\Lie^n(Y)\Theta^\nabla_h=\d\zeta_Y$.
 \item
 If $\Lie^{n-1}(Y)\inn(Y)\Omega^\nabla_h=\d\xi_Y$,
 in an open set $U\subset J^{1*}E$, then
 $$
 \Lie^n(Y)\Theta^\nabla_h=
 \d (\Lie^{n-1}(Y)\inn(Y)\Theta^\nabla_h-\xi_Y)=\d\zeta_Y
 \quad \mbox{\rm (in $U$)}
 $$
 \een
 \end{prop}
 \proof
 \ben
 \item
 The first item is immediate since
 $\d\Lie^n(Y)\Theta^\nabla_h=\Lie^n(Y)\d\Theta^\nabla_h=0$.
 \item
 For the second item we have
 \beann
 \Lie^n(Y)\Theta^\nabla_h &=&
 \Lie^{n-1}(Y)\Lie(Y)\Theta^\nabla_h=
 \Lie^{n-1}(Y)(\d\inn(Y)\Theta^\nabla_h+\inn(Y)\d\Theta^\nabla_h)
 \\ &=&
 \d\Lie^{n-1}(Y)\inn(Y)\Theta^\nabla_h+\Lie^{n-1}(Y)\inn(Y)\d\Theta^\nabla_h
 \\ &=&
 \d\Lie^{n-1}(Y)\inn(Y)\Theta^\nabla_h-\d\xi_Y=
 \d (\Lie^{n-1}(Y)\inn(Y)\Theta^\nabla_h-\xi_Y)
 \eeann
 Hence we can write $\xi_Y=\Lie^{n-1}(Y)\inn(Y)\Theta^\nabla_h-\zeta_Y$.
 \een
 \qed

 Then, theorem \ref{Nth}
 can be generalized for including higher-order Cartan-Noether symmetries:

 \begin{teor}
 {\rm (Noether):}
 If $Y\in\vf (J^{1*}E)$ is a Cartan-Noether symmetry of order $n$
 of a Hamiltonian system $\hs$,
 with $\Lie^{n-1}(Y)\inn(Y)\Omega^\nabla_h=\d\xi_Y$. Then,
 for every HDW-multivector field
 $X_{\cal H}\in\vf^m(J^{1*}E)$, we have that
 $$
 \Lie(X_{\cal H})\xi_Y=0
 $$
 that is, the $(m-1)$-form
 $\xi_Y$ associated with $Y$ is a first integral of $\hs$.
 \label{Nthgen}
 \end{teor}
 \proof
 If $Y\in\vf (J^{1*}E)$ is a Cartan-Noether symmetry then
 it is a general symmetry, and then
 $[Y,X_{\cal H}]={\cal Z}\in\ker\,\Omega^\nabla_h$.
 Therefore
 \beann
 \Lie(X_{\cal H})\xi_Y &=&
 (-1)^{m+1}\inn(X_{\cal H})\d\xi_Y=
 (-1)^{m+1}\inn(X_{\cal H})\Lie^{n-1}(Y)\inn(Y)\Omega^\nabla_h
 \\ &=&
 (-1)^{m+1}\inn(X_{\cal H})\Lie(Y)\Lie^{n-2}(Y)
  \inn(Y)\Omega^\nabla_h
  \\ &=&
 \Lie(Y)\inn(X_{\cal H})\Lie^{n-2}(Y)\inn(Y)\Omega^\nabla_h
 -\inn([Y,X_{\cal H}])\Lie^{n-2}(Y)\inn(Y)\Omega^\nabla_h
  \\ &=&
 (\Lie(Y)\inn(X_{\cal H})-\inn({\cal Z}))
 \Lie^{n-2}(Y)\inn(Y)\Omega^\nabla_h
 \eeann
 and repeating the reasoning $n-2$ times we will arrive at the
 result
 $$
 \Lie(X_{\cal H})\xi_Y=
 (\Lie(Y)\inn(X_{\cal H})-\inn({\cal Z}))^{n-1}
 \inn(Y)\Omega^\nabla_h=0
 $$
 since $\inn(X_{\cal H})\inn(Y)\Omega^\nabla_h=0$
 and $\inn({\cal Z})\inn(Y)\Omega^\nabla_h=0$.
 \qed

 The study of symmetries of Hamiltonian multisymplectic systems,
 is, of course, a topic of great interest. The general problem of
 a group of symmetries acting on a
 multisymplectic manifold and the subsequent theory of reduction
 has been analyzed in \cite{Hr-99a} and \cite{Hr-99b}.

 \section{Restricted Hamiltonian systems}
 \protect\label{hfjfbis}

 There are many interesting cases in Field theories
 where the Hamiltonian field equations are established
 not in the whole multimomentum phase space $J^{1*}E$,
 but rather in a submanifold
 ${\rm j}_0\colon P\hookrightarrow J^{1*}E$,
 such that $P$ is a fiber bundle over $E$ (and $M$), and
 the corresponding projections
 $\tau^1_0\colon P\to E$ and $\bar\tau^1_0\colon P\to M$
 satisfy that $\tau^1\circ{\rm j}_0=\tau^1_0$ and
 $\bar\tau^1\circ{\rm j}_0=\bar\tau^1_0$
 In that case we will say that $\hso$ is a
 {\sl restricted Hamiltonian system}, where
 $\Omega^0_h:={\rm j}_0^*\Omega^\nabla_h$.

 Now we can pose a variational principle in the same way
 as for the Hamiltonian system $\hs$,
 (but with $P$ instead of $J^{1*}E$):
 the states of the field are the sections of
 $\bar\tau^1_0$ which are critical for the functional
 ${\bf H}_0\colon\Gamma_c(M,P)\to\Real$
 defined by ${\bf H}_0(\psi_0):=\int_M\psi_0^*\Theta^0_h$,
 for every $\psi_0\in\Gamma_c(M,P)$.
 These critical sections will be characterized by the condition
 (analogous to (\ref{eqn0}))
 $$
 \psi_0^*\inn (X_0)\Omega^0_h = 0 \quad ,\quad
 \mbox{for every} \ X_0\in\vf (P)
 $$
 Hence, considering multivector fields, connections
 and jet fields in $P$ instead of $J^{1*}E$, we have:

 \begin{prop}
 Let $\hso$ be a restricted Hamiltonian system.
 The critical section of
 the above variational principle are sections
 $\psi_0\in\Gamma_c(M,P)$ satisfying the
 following equivalent conditions:
 \ben
\item
They are the integral sections of an integrable jet field ${\cal
Y}^0_{\cal H}\colon P\to J^1P$ satisfying that $\inn ({\cal
Y}^0_{\cal H})\Omega^0_h=0$.
\item
They are the integral sections of an integrable connection
$\nabla^0_{\cal H}$ satisfying that $\inn (\nabla^0_{\cal
H})\Omega^0_h=(m-1)\Omega^0_h$.
\item
They are the integral sections of a class of integrable and
$\bar\tau_0^1$-transverse multivector fields $\{ X^0_{\cal
H}\}\subset\vf^m(P)$ such that $\inn (X^0_{\cal H})\Omega^0_h=0$,
for every $X^0_{\cal H}\in\{ X^0_{\cal H}\}$. \een
\end{prop}
 \proof
 The proof is like in Theorem \ref{hameq}.
 \qed

 Note that the form $\Omega^0_h$ is $m$-degenerate but, in
 general, a $\bar\tau_0^1$-transverse and locally decomposable
 multivector field $X^0_{\cal H}\in\vf^m(P)$ such that $\inn
 (X^0_{\cal H})\Omega^0_h=0$, does not necessarily exist.
 Furthermore, the existence of multivector fields of this kind does
 not imply their integrability. Nevertheless, it is possible for
 these integrable multivector fields to exist on a submanifold of
 $P$. So we can state the following problem: to look for a
 submanifold $S\hookrightarrow P$ where integrable
 HDW-multivector fields $X^0_{\cal H}\in\vf^m(P)$ exist; and then their
 integral sections are contained in $S$.

As a first step, we do not consider the integrability condition.
The procedure is algorithmic (from now on we suppose that all the
multivector fields are locally decomposable): \bit
\item
First, let $S_1$ be the set of points of $P$ where HDW-multivector
fields do exist
 $$ S_1:=\{ \tilde y\in P \ ;\
\exists X^0_{\cal H}\in\vf^m(P)\ \mbox{such that} \left\{
\begin{array}{c}
(\inn(X^0_{\cal H})\Omega^0_h)(\tilde y)=0
\\
(\inn(X^0_{\cal H})(\bar\tau^{1*}_0\omega))(\tilde y)=1
\end{array}
\right\} \} $$ We assume that $S_1$ is a non-empty (closed)
submanifold of $P$.

This is the {\sl compatibility condition}.
\item
Now, denote by $\vf^m_{HDW}(P,S_1)$ the set of multivector fields
in $P$ which are HDW-multivector fields on $S_1$. Let $X^0_{\cal
H}\colon S_1\to\Lambda^m\Tan P\vert_{S_1}$ be in $\vf^m_{\cal
H}(P,S_1)$. If, in addition, $X^0_{\cal H}\colon
S_1\to\Lambda^m\Tan S_1$; that is, $X^0_{\cal H}\in\vf^m(S_1)$,
then we say that $X^0_{\cal}$ is a solution on $S_1$.
Nevertheless, this last condition is not assured except perhaps in
a set of points $S_2\subset S_1\subset P$, which we will assume to
be a (closed) submanifold, and which is defined by
 $$
 S_2:=\{\tilde y\in S_1 \ ;\
 \exists X^0_{\cal H}\in\vf^m_{HDW}(P,S_1)\ \mbox{such that}\
 X^0_{\cal H}(\tilde y)\in\Lambda^m\Tan_{\tilde y}S_1\}
 $$
 This is the so-called {\sl consistency} or
 {\sl tangency condition}.
\item
 In this way, a sequence of (closed) submanifolds,
 $\ldots \subset S_i\subset\ldots \subset S_1\subset P$,
 is assumed to be obtained, each one of them being defined as
 $$ S_i:=\{ \tilde y\in S_{i-1} \
 ;\ \exists X^0_{\cal H}\in\vf^m_{HDW}(P,S_{i-1})\ \mbox{such
 that}\ X^0_{\cal H}(\tilde y)\in\Lambda^m\Tan_{\tilde y}S_{i-1}\}
 $$
\item
There are two possible options for the final step of this
algorithm, namely: \ben
\item
The algorithm ends by giving a submanifold $S_f\hookrightarrow P$,
with $\dim\, S_f\geq m$, (where \dst S_f=\bigcap_{i\geq 1}S_i\) )
and HDW-multivector fields $X^0_{\cal H}\in\vf^m(S_f)$. $S_f$ is
then called the {\sl final constraint submanifold}.
\item
The algorithm ends by giving a submanifold $S_f$ with $\dim\,
S_f<m$, or the empty set. Then there is no HDW-multivector fields
$X^0_{\cal H}\in\vf^m(S_f)$.
 \een \eit
  This procedure is called the {\sl constraint algorithm}.

 The local treatment of this case shows significative differences
 to the general one. We again have the system of equations for the
 coefficients $G^\mu_{A\nu}$. As we have stated, this system is not
 compatible in general, and $S_1$ is the closed submanifold where
 it is compatible. Then, there are HDW-multivector fields on
 $S_1$, but the number of arbitrary functions on which they depend
 is not the same as in the general case, since it depends on the
 dimension of $S_1$. Now the tangency condition must be analyzed in
 the usual way. Finally, the question of integrability must be
 considered. To this purpose similar considerations as above must
 be made for the submanifold $S_f$ instead of $J^{1*}E$.

 Some of the problems considered in this and the above section have
 been treated in an equivalent way, but using Ehresmann
 connections, in \cite{LMM-95} and \cite{LMM-96}.

 As a final remark, concerning to the study of symmetries for
 restricted Hamiltonian systems, results like those
 discussed in sections \ref{ntmvf} and  \ref{ntmvfbis}
 would be applicable,in general,
 to this situation, but for the subbundle
 $S_f\to M$, and taking as symmetries vector fields
 $Y\in\vf (J^{1*}E$ which are tangent to $S_f$.

 \section{Hamiltonian formalism for Lagrangian systems}
 \protect\label{vp}

 From the Lagrangian point of view, a Classical Field theory is
described by its {\sl configuration bundle} $\pi\colon E\to M$,
and a {\sl Lagrangian density} which is a $\bar\pi^1$-semibasic
$m$-form in $J^1E$. A Lagrangian density is usually written as
$\Lag =\lag \bar\pi^{1*}\omega$, where $\lag\in\Cinfty (J^1E)$ is
the {\sl Lagrangian function} associated with $\Lag$ and $\omega$.
Then a {\sl Lagrangian system} is a couple $\ls$. The {\sl
Poincar\'e-Cartan $m$ and $(m+1)$-forms} associated with the
Lagrangian density $\Lag$ are defined using the {\sl vertical
endomorphism} ${\cal V}$ of the bundle $J^1E$ \cite{EMR-96},
\cite{Gc-73}:
 $$ \Theta_{\Lag}:=\inn({\cal
 V})\Lag+\Lag\equiv\theta_{\Lag}+\Lag\in\df^{m}(J^1E) \quad ;\quad
 \Omega_{\Lag}:= -\d\Theta_{\Lag}\in\df^{m+1}(J^1E)
 $$
 In a natural chart in $J^1E$ we have
 \beann
 \Theta_{\Lag}&=&\derpar{\lag}{v^A_\mu}\d
y^A\wedge\d^{m-1}x_\mu - \left(\derpar{\lag}{v^A_\mu}v^A_\mu
-\lag\right)\d^mx
\\
\Omega_{\Lag}&=& -\frac{\partial^2\lag}{\partial v^B_\nu\partial
v^A_\mu} \d v^B_\nu\wedge\d y^A\wedge\d^{m-1}x_\mu
-\frac{\partial^2\lag}{\partial y^B\partial v^A_\mu}\d y^B\wedge
\d y^A\wedge\d^{m-1}x_\mu + \\ & & \frac{\partial^2\lag}{\partial
v^B_\nu\partial v^A_\mu}v^A_\mu \d v^B_\nu\wedge\d^mx  +
\left(\frac{\partial^2\lag}{\partial y^B\partial v^A_\mu}v^A_\mu
-\derpar{\lag}{y^B}+ \frac{\partial^2\lag}{\partial x^\mu\partial
v^B_\mu} \right)\d y^B\wedge\d^mx \eeann
 The Lagrangian system is
{\sl regular} if $\Omega_{\Lag}$ is $1$-nondegenerate and, as a
consequence, $(J^1E,\Omega_{\Lag})$ is a multisymplectic manifold
\cite{CIL-98}. Elsewhere the system is {\sl non-regular} or {\sl
singular}.  The regularity condition is equivalent to demanding
that \dst det\left(\frac{\partial^2\lag}{\partial v^A_\mu\partial
v^B_\nu}(\bar y)\right)\not= 0\) , for every $\bar y\in J^1E$.
(For more details see, for instance, \cite{BSF-88}, \cite{CCI-91},
\cite{EMR-96}, \cite{Gc-73}, \cite{GMS-97}, \cite{GS-73},
\cite{Sd-95}, \cite{Sa-89}).

 As for Hamiltonian systems, a variational problem can be posed
 for a Lagrangian system, which is called the {\sl Hamilton principle}
 of the Lagrangian formalism: the states of the field are the
 (compact-supported) sections of $\pi$
 which are critical for the functional
 ${\bf L}\colon\Gamma_c(M,E)\to\Real$ defined by
 ${\bf L}(\phi):=\int_M(j^1\phi)^*\Lag$,
 for every $\phi\in\Gamma_c(M,E)$.
 These (compact-supported) critical sections are
 characterized by the condition
 $$
 (j^1\phi)^*\inn (X)\Omega_{\Lag} = 0 \quad ,\quad
 \mbox{for every} \ X\in\vf (J^1E)
 $$
 which, in a natural system of coordinates in $J^1E$,
 is equivalent to demanding that
 $\phi$ satisfy the {\sl Euler-Lagrange equations}:
 \dst \derpar{\lag}{y^A}\Big\vert_{j^1\phi}-
 \derpar{}{x^\mu}\derpar{\lag}{v_\mu^A}\Big\vert_{j^1\phi} = 0 \) .
 Then
 \cite{EMR-96}, \cite{EMR-98}, \cite{LMM-95}, \cite{Sa-89}:

 \begin{teor}
 The critical sections of the Hamilton principle are
 canonical liftings, $j^1\phi\colon M\to J^1E$, of sections
 $\phi\colon M\to E$, which satisfy any one of the following conditions:
 \ben
 \item
 They are the integral sections of an holonomic jet field ${\cal
 Y}_{\Lag}\colon J^1E\to J^1J^1E$ such that $\inn ({\cal
 Y}_{\Lag})\Omega_{\Lag}=0$.
 \item
 They are the integral sections of an holonomic connection
 $\nabla_{\Lag}$ such that $\inn
 (\nabla_{\Lag})\Omega_{\Lag}=(m-1)\Omega_{\Lag}$.
 \item
 They are the integral sections of a class of holonomic multivector
 fields $\{ X_{\Lag}\}\subset\vf^m(J^1E)$ such that $\inn
 (X_{\Lag})\Omega_{\Lag}=0$, for every $X_{\Lag}\in\{ X_{\Lag}\}$.
 \een \label{importantlag}
 \end{teor}

 $X_{\Lag}\in\vf^m(J^1E)$ is an {\sl Euler-Lagrange
 multivector field} for $\Lag$ if it is semi-holonomic and is a
 solution of the  equation $\inn (X_{\Lag})\Omega_{\Lag}=0$. (The
 same terminology is also used for jet fields and connections).
 Then, using this theorem, it can be proved that
 \cite{EMR-98}, \cite{LMM-95}:
  \bit
\item
If $\ls$ is a regular Lagrangian system, then there exist classes
of Euler-Lagrange multivector fields for $\Lag$. In a local system
these multivector fields depend on $N(m^2-1)$ arbitrary functions,
and they are not integrable necessarily, except perhaps on a
submanifold $I\hookrightarrow J^1E$; such that the integral
sections are in $I$.
\item
For singular Lagrangian systems, the existence of Euler-Lagrange
multivector fields is not assured, except perhaps on some
submanifold $S\hookrightarrow J^1E$. Furthermore, locally
decomposable and $\bar\pi^1$-transverse multivector fields, which
are solutions of the field equations, can exist (in general, on
some submanifold of $J^1E$), but none of them are semi-holonomic
(at any point of this submanifold). As in the regular case,
although Euler-Lagrange multivector fields exist on some
submanifold $S$, their integrability is not assured, except
perhaps on another smaller submanifold $I\hookrightarrow S$. \eit

 The Lagrangian and Hamiltonian formalisms are related by means
 of the corresponding {\sl Legendre map}
 $F\Lag\colon J^1E\to J^{1*}E$. In order to define it, first we
 introduce the {\sl extended Legendre map}
 $\widetilde{{\cal F}\Lag}\colon J^1E\to {\cal M}\pi$
 in the following way \cite{LMM-96}:
  $$
 (\widetilde{{\cal F}\Lag}\bar y))(\moment{Z}{1}{m}):=
 (\Theta_{\Lag})_{\bar y}(\moment{\bar Z}{1}{m})
 $$
 where $\moment{Z}{1}{m}\in\Tan_{\pi^1(\bar y)}E$, and
 $\moment{\bar Z}{1}{m}\in\Tan_{\bar y}J^1E$ are such that
 $\Tan_{\bar y}\pi^1\bar Z_\mu=Z_\mu$.
 ($\widetilde{{\cal F}\Lag}$ can also be defined as the
  ``first order  vertical Taylor approximation to
 $\lag$'' \cite{CCI-91}, \cite{GIMMSY-mm}).
 Hence, using the natural projection
 $\mu \colon {\cal M}\pi=\Lambda_1^m\Tan^*E \to
 \Lambda_1^m\Tan^*E/\Lambda_0^m\Tan^*E=J^{1*}E$,
 we define $F\Lag :=\mu\circ\widetilde{{\cal F}\Lag}$.
 Its local expression is
 $$
 F\Lag^*x^\mu = x^\mu \quad\ , \ \quad
 F\Lag^*y^A = y^A \quad\  , \quad
 F\Lag^*p_A^\mu =\derpar{\lag}{v^A_\mu}
 $$

 \begin{definition}
 Let $\ls$ be a Lagrangian system. \ben
 \item
 $\ls$ is a {\rm regular} or {\rm non-degenerate}
 Lagrangian system if $F\Lag$ is a local diffeomorphism.
 Elsewhere $\ls$ is a {\rm singular} or
 {\rm degenerate} Lagrangian system
 (This definition is equivalent to that given above).

 As a particular case, $\ls$ is a {\rm hyper-regular}
 Lagrangian system if $F\Lag$ is a global diffeomorphism.
 \item
 A singular Lagrangian system $\ls$ is {\rm
 almost-regular} if:
  \ben
 \item
 $P:={\rm F}\Lag (J^1E)$ is a closed submanifold of $J^{1*}E$.

 (We will denote the natural imbedding by
  ${\rm j}_0\colon P\hookrightarrow J^{1*}E$).
\item
 $F\Lag$ is a submersion onto its image.
\item
 For every $\bar y\in J^1E$, the fibers $F\Lag^{-1}(F\Lag (\bar y))$
 are connected submanifolds of $J^1E$.
 \een
 \een
 \end{definition}

 It can be proved \cite{CCI-91}, \cite{LMM-96}, that
 if $\ls$ is a hyper-regular Lagrangian system, then
 $\widetilde{{\cal F}\Lag}(J^1E)$ is a
 1-codimensional imbedded submanifold of ${\cal M}\pi$, which is
 transverse to the projection $\mu$, and is diffeomorphic to
 $J^{1*}E$. This diffeomorphism is
 ${\rm h}:=\widetilde{{\cal F}\Lag}\circ F\Lag^{-1}$
 (which is just $\mu^{-1}$, when $\mu$ is
 restricted to $\widetilde{{\cal F}\Lag}(J^1E)$),
 and it is a Hamiltonian section.
 Thus we can construct the Hamilton-Cartan forms by making
 $\Theta_h={\rm h}^*\Theta$ and $\Omega_h={\rm h}^*\Omega$.
 Then the couple $\hsl$ is said to be the {\sl Hamiltonian system}
 associated with the hyper-regular Lagrangian system $\ls$.
 Locally, this Hamiltonian section is specified by the
 local Hamiltonian function
 $H=p^\mu_A F\Lag^{-1^*}v_\mu^A-F\Lag^{-1^*}\lag$,
 then the local expressions of these Hamilton-Cartan forms are
 (\ref{omegaHlocal}), and the (non-covariant) expression
 of the Hamiltonian equations are (\ref{hdw}).
 Of course
 $F\Lag^*\Theta_h=\Theta_{\Lag}$ and
 $F\Lag^*\Omega_h=\Omega_{\Lag}$.

 This construction can also be made as follows:
 given a connection $\nabla$ in the bundle $\pi\colon E\to M$, let
 $j_\nabla\colon J^{1*}E\to{\cal M}\pi$ be the associated linear
 section, and $\Theta^\nabla=j_\nabla^*\Theta$. Then we can define
 a unique Hamiltonian density ${\cal H}^\nabla$
 in two different but equivalent ways:
 by making the difference $j_\nabla-{\rm h}$, or
 by making $(F\Lag^{-1})^*\del$, where $\del$ is
 the {\sl density of Lagrangian energy} of the Lagrangian
 formalism constructed using the connection $\nabla$.
 In any case, the form
 $\Theta_h=\Theta^\nabla-{\cal H}^\nabla$,
 and hence $\Omega_h$, are the same as above
 (see \cite{EMR-99b}).

 If $\ls$ is an almost-regular Lagrangian system,
 then a {\sl restricted Hamiltonian system}
 $\hso$ can be associated in a similar way
 \cite{EMR-99b}, \cite{LMM-96}.

 One expects both the Lagrangian and
Hamiltonian formalism to be equivalent. As in mechanics, the
standard way of showing this equivalence consists in using the
Legendre map. First we can lift sections of $\pi$ from $E$ to
$J^{1*}E$, as follows:

\begin{definition}
Let $\ls$ be a hyper-regular Lagrangian system, $F\Lag$ the
induced Legendre transformation, $\phi\colon M\to E$ a section of
$\pi$ and $j^1\phi\colon M\to J^1E$ its canonical prolongation to
$J^1E$. The {\rm Lagrangian prolongation} of $\phi$ to $J^{1*}E$
 is the section
 $$
 j^{1*}\phi:=F\Lag\circ j^1\phi\colon M\to J^{1*}E
 $$
 (If $\ls$ is an almost-regular Lagrangian system, the
 {\sl Lagrangian prolongation} of a section
 $\phi\colon M\to E$ to $P$ is
 $j_0^{1*}\phi:={\rm F}\Lag_0\circ j^1\phi\colon M\to P$).
 \end{definition}

\begin{teor}
{\rm (Equivalence theorem for sections)} Let $\ls$ and $\hsl$ be
the Lagrangian and Hamiltonian descriptions of a hyper-regular
system.

If a section $\phi\in\Gamma_c(M,E)$ is a solution of the
Lagrangian variational problem (Hamilton principle) then the
section $\psi\equiv j^{1*}\phi:=F\Lag\circ
j^1\phi\in\Gamma_c(M,J^{1*}E)$ is a solution of the Hamiltonian
variational problem (Hamilton-Jacobi principle).

Conversely, if $\psi\in\Gamma_c(M,J^{1*}E)$ is a solution of the
Hamiltonian variational problem, then the section
$\phi\equiv\tau^1\circ\psi\in\Gamma_c(M,E)$ is a solution of the
Lagrangian variational problem. \label{equiv1}
\end{teor}
\proof Bearing in mind the diagram \beq
\begin{array}{ccc}
J^1E &
\begin{picture}(135,10)(0,0)
\put(63,6){\mbox{$F\Lag$}} \put(0,3){\vector(1,0){135}}
\end{picture}
& J^{1*}E
\\ &
\begin{picture}(135,100)(0,0)
\put(34,82){\mbox{$\pi^1$}} \put(90,82){\mbox{$\tau^1$}}
\put(30,55){\mbox{$j^1\phi$}} \put(96,55){\mbox{$\psi$}}
\put(75,30){\mbox{$\pi$}} \put(55,30){\mbox{$\phi$}}
\put(63,55){\mbox{$E$}} \put(65,0){\mbox{$M$}}
\put(0,100){\vector(3,-2){55}} \put(135,100){\vector(-3,-2){55}}
\put(55,13){\vector(-2,3){55}} \put(80,13){\vector(2,3){55}}
\put(65,13){\vector(0,1){30}} \put(71,43){\vector(0,-1){30}}
\end{picture} &
\end{array}
\label{dia} \eeq If $\phi$ is a solution of the Lagrangian
variational problem then $(j^1\phi)^*\inn (X)\Omega_{\Lag}=0$, for
every $X\in\vf (J^1E)$ (Theorem \ref{importantlag}); therefore, as
$F\Lag$ is a local diffeomorphism, \beann 0&=&(j^1\phi)^*\inn
(X)\Omega_{\Lag}= (j^1\phi)^*\inn (X)(F\Lag^*\Omega_h)
\\ &=&
(j^1\phi)^*F\Lag^*(\inn (F\Lag_*^{-1}X)\Omega_h)= (F\Lag\circ
j^1\phi)^*\inn (X')\Omega_h) \eeann which holds for every
$X'\in\vf (J^{1*}E)$ and thus, by (\ref{eqn0}),
 $\psi\equiv F\Lag\circ j^1\phi$
 is a solution of the Hamiltonian variational
 problem. (This proof also holds for the almost-regular case).

Conversely, let $\psi\in\Gamma_c(M,J^{1*}E)$ be a solution of the
Hamiltonian variational problem. Reversing the above reasoning we
obtain that $({\rm F}\Lag^{-1}\circ\psi)^*\inn
(X)\Omega_{\Lag}=0$, for every $X\in\vf (J^1E)$, and hence
$\sigma\equiv F\Lag^{-1}\circ\psi\in\Gamma_c(M,J^1E)$ is a
critical section for the Lagrangian variational problem. Then, as
we are in the hyper-regular case, $\sigma$ must be an holonomic
section, $\sigma =j^1\phi$ \cite{EMR-98}, \cite{LMM-95},
\cite{Sa-89}, and since (\ref{dia}) is a commutative diagram,
$\phi =\tau^1\circ\psi\in\Gamma_c(M,E)$. \qed

 Observe that every section $\psi\colon M\to J^{1*}E$ which is
 solution of the Hamilton-Jacobi variational principle is
 necessarily a Lagrangian prolongation of a section
 $\phi\colon M\to E$.

 \begin{teor}
 Let $\ls$ and $\hsl$ be the Lagrangian and Hamiltonian
 descriptions of a hyper-regular system.
 \ben
 \item
 {\rm (Equivalence theorem for jet fields and connections)} Let
 ${\cal Y}_{\Lag}$ and ${\cal Y}_{\cal H}$ be the jet fields
 solution of the Lagrangian and Hamiltonian field equations
 respectively. Then
 $$
 j^1F\Lag\circ{\cal Y}_{\Lag}={\cal Y}_{\cal H}\circ F\Lag
 $$
 (we say that the jet fields ${\cal Y}_{\Lag}$ and
 ${\cal Y}_{\cal H}$ are $F\Lag$-related). As a consequence, their
associated connection forms, $\nabla_{\Lag}$ and $\nabla_{\cal H}$
respectively, are $F\Lag$-related too.
\item
{\rm (Equivalence theorem for multivector fields)} Let
$X_{\Lag}\in\vf^m(J^1E)$ and $X_{\cal H}\in\vf^m(J^{1*}E)$ be
multivector fields solution of the Lagrangian and Hamiltonian
field equations respectively. Then $$ \Lambda^m\Tan F\Lag\circ
X_{\Lag}=fX_{\cal H}\circ F\Lag $$ for some $f\in\Cinfty
(J^{1*}E)$ (we say that the classes $\{ X_{\Lag}\}$ and $\{
 X_{\cal H}\}$ are $F\Lag$-related). \een
That is, we have the following (commutative) diagrams:
   $$
\begin{array}{ccc}
\Lambda^m\Tan J^1E & $\rightarrowfill$ & \Lambda^m\Tan J^{1*}E
\\
& \Lambda^m\Tan F\Lag &
\\
X_{\Lag}\ \Big\uparrow & &\Big\uparrow\ X_{\cal H}
\\
& F\Lag &
\\
J^1E & $\rightarrowfill$ & J^{1*}E
\end{array}
\qquad ;\qquad
\begin{array}{ccc}
J^1J^1E & $\rightarrowfill$ & J^1(J^{1*}E)
\\
& j^1F\Lag &
\\
{\cal Y}_{\Lag}\ \Big\uparrow & &\Big\uparrow\ {\cal Y}_{\cal H}
\\
& F\Lag &
\\
J^1E & $\rightarrowfill$ & J^{1*}E
\end{array}
$$
 \label{equiv2}
\end{teor}
\proof
 The first item is a consequence of Theorem \ref{equiv1},
 since the critical sections solutions of the Lagrangian and
 Hamiltonian variational problems (which are $F\Lag$-related) are
 the integral sections of the jet fields ${\cal Y}_{\Lag}$ and
 ${\cal Y}_{\cal H}$, respectively (see also \cite{LMM-96}).

 The second item is an immediate consequence of the first one and
 the equivalence between orientable and integrable jet fields and
 classes of non-vanishing, locally decomposable, transverse and
 integrable multivector fields.
 \qed

 \section{Example}

 (See also the reference  \cite{Sd-95}).

 Most of the (quadratic) Lagrangian systems in field theories can
 be modeled as follows: $\pi\colon E\to M$ is a trivial bundle
 (usually $E=M\times\Real^N$) and then $\pi^1\colon J^1E\to E$ is a
 vector bundle. $g$ is a metric in this vector bundle,
 $\gamma$ is a connection of the projection $\pi^1$, and
 $f\in\Cinfty(E)$ is a potential function. Then the Lagrangian
 function is
 $$
 \lag (\bar y) =
 \frac{1}{2}g(\bar y-\gamma(\pi^1(\bar y)),\bar y-
 \gamma(\pi^1(\bar y)))+({\pi^1}^*f)(\bar y)
 \qquad \mbox{(for $\bar y\in J^1E$)}
 $$
 and in natural coordinates takes the form
 \cite{Sd-95}
 $$
 \lag =\frac{1}{2}a_{AB}^{\mu\nu}(y)
 (v^A_\mu-\gamma^A_\mu(x))(v^B_\nu-\gamma^B_\nu(x))+f(y)
 $$
 For simplicity, we consider a model where the matrix of the
 coefficients $a_{AB}^{\mu\nu}$ is regular and symmetric at every
 point (that is, $a^{\mu\nu}_{AB}(y)=a^{\nu\mu}_{BA}(y)$). This
 fact is equivalent to the non-degeneracy of the metric $g$.
 The Legendre map associated with this Lagrangian system is given by
 $$
 F\Lag^*x^\mu = x^\mu \quad\ , \ \quad
 F\Lag^*y^A = y^A \quad\  , \quad
 F\Lag^*p_A^\mu =a_{AB}^{\mu\nu}(y)(v^B_\nu-\gamma^B_\nu(x))
 $$
 and the local expression of the Hamilton-Cartan $(m+1)$-form is
 (\ref{omegaHlocal}), where the local Hamiltonian function is
 $$
 H=\frac{1}{2}\tilde a^{AB}_{\mu\nu}(y)p_A^\mu p_B^\nu-f(y)
 $$
 (here $\tilde a^{AB}_{\mu\nu}$ denote the coefficients of the
 inverse matrix  of $(a_{AB}^{\mu\nu})$). Hence
 \beann
 \Theta^\nabla_h &=& p_A^\mu\d y^A\wedge\d^{m-1}x_\mu-
 \left(\frac{1}{2}\tilde a^{AB}_{\mu\nu}(y)p_A^\mu p_B^\nu
 -f(y)\right)\d^mx
 \\
 \Omega^\nabla_h &=& -\d p_A^\mu\wedge\d y^A\wedge\d^{m-1}x_\mu+
 \d\left(\frac{1}{2}\tilde a^{AB}_{\mu\nu}(y)p_A^\mu p_B^\nu
 -f(y)\right)\wedge\d^mx
 \eeann
 and it is a multisymplectic form. Then,
 taking (\ref{locmvf}) as the local expression for
 representatives of the corresponding classes of
 HDW-multivector fields $\{ X_{\cal H}\}\subset\vf^m_{HDW}(J^{1*}E)$,
 their components $F^A_\mu$ are
 $$
 F^A_\mu=\derpar{H}{p^\mu_A}=\tilde a^{AB}_{\mu\nu}(y)p_B^\nu
 $$
 and $G^\mu_{A\nu}$ are related by the equations
 \beq
 G^\rho_{A\rho}=-\derpar{H}{y^A}=
 -\frac{1}{2}\derpar{\tilde a^{CB}_{\mu\nu}}{y^A}p_C^\mu p_B^\nu+
 \derpar{f}{y^A}
 \label{eqsGexamp}
 \eeq
 This system allows us to isolate $N$ of
 these components as functions of the remaining $N(m^2-1)$; and
 then it determines a family of (classes of)
 HDW-multivector fields. In order to obtain an integrable class, the
 condition of integrability ${\cal R}=0$ (where ${\cal R}$ is the
 curvature of the associated connection) must hold; that is,
 equations (\ref{curvcero1})
 and (\ref{curvcero2}) must be added to the last system.

 As a simpler case, we consider that the matrix of coefficients is
 $\tilde a_{\mu\nu}^{AB}(y)=\delta^{AB}\delta_{\mu\nu}$,
 (that is, we take
 an orthonormal frame for the metric $g$), then we have that
 \dst H =\frac{1}{2}\delta^{AB}\delta_ {\mu\nu}p_A^\mu p_B^\nu-f(y)\) .
 In this case, equations (\ref{eqsGexamp}) reduce to
 $$
 G^\rho_{A\rho}=\derpar{f}{y^A}
  $$
 From this system we can isolate $N$ of the coefficients $G^\mu_{A\nu}$;
 for instance, if $\mu,\nu=0,1,\ldots ,m-1$, those for which
 $\mu=\nu=0$: Thus
 $$
 G^0_{A0}=\derpar{f}{y^A}-\sum_{\mu=1}^{m-1}\delta^{AB}G^\mu_{B\mu}
 $$
 Therefore the HDW-multivector fields are
 \beann
 X_{\cal H}&=&
 \bigwedge_{\mu=0}^{m-1}\left(\derpar{}{x^\mu}+
 \delta^{AB}\delta_{\mu\nu}p_B^\nu\derpar{}{y^A}+
 \delta_\mu^0\left[\derpar{f}{y^A}-
 \sum_{\nu=1}^{m-1}\delta^{AB}G^\mu_{B\mu}\right]\derpar{}{p^0_A}\right.
 \\ & &
 \left. + \sum_{\mu=\eta\not=0}G^\mu_{B\eta}\derpar{}{p^\eta_B}+
 \sum_{\mu\not=\eta}G^\mu_{C\eta}\derpar{}{p^\eta_C}\right)
 \eeann
 Now, if we look for integrable Euler-Lagrange multivector
 fields, the integrability conditions (\ref{curvcero1})
 and (\ref{curvcero2}) must be imposed.

 The Lagrangian formalism for this model (using multivector fields)
 has been studied in \cite{EMR-98}. Then, the
 corresponding (semi-holonomic) Euler-Lagrange multivector fields $X_{\Lag}$
 given there by
 \beann
 X_{\Lag}&=&
 \bigwedge_{\mu=0}^{m-1}\left(\derpar{}{x^\mu}+v_\mu^A\derpar{}{y^A}+
 \delta_{0\mu}\delta^{AC}\left[\derpar{f}{y^C}-
 \sum_{\nu=1}^{m-1}\delta_{CD}\bar G^D_{\nu\nu}\right]\derpar{}{v_0^A}
 \right.
 \\ & & \left. +
 \sum_{\mu=\eta\not=0}\bar G_{\mu\eta}^B\derpar{}{v_\eta^B}+
 \sum_{\mu\not=\eta}\bar G_{\mu\eta}^C\derpar{}{v_\eta^C}\right)
 \eeann
 can be compared with the HDW-multivector fields
 here obtained, observing that, in fact, they are related
 as stated in the second item of Theorem \ref{equiv2}.

 As a final remark, we can obtain some typical first integrals,
 by applying Noether's theorem.
 As infinitesimal generators of symmetries we take
 the following $\pi$-projectable vector fields in $E$
  $$
 Z_\mu = \derpar{}{x^\mu} \quad , \quad
 Z_{\mu\nu}=x^\mu\derpar{}{x^\nu}-x^\nu\derpar{}{x^\mu}
 $$
 (they are isometries of the metric $g$ and symmetries of the potential
 function $f$, which generate space-time translations and rotations),
 and whose canonical liftings to $J^{1*}E$ are the vector fields
 $$
 Y_\mu =\derpar{}{x^\mu}
 \quad , \quad
 Y_{\mu\nu} =x^\mu\derpar{}{x^\nu}-x^\nu\derpar{}{x^\mu}+
 p^\nu_A\derpar{}{p^\mu_A}-p^\mu_A\derpar{}{p^\nu_A}
 $$
 In fact, they are Cartan-Noether symmetries satisfying that
 $\Lie(Y_\mu)\Theta^\nabla_h=0$ and
 $\Lie(Y_{\mu\nu})\Theta^\nabla_h=0$,
 and their corresponding associated first integrals are then
 \beann
 \xi_{Y_\mu}&=&\inn(Y_\mu)\Theta^\nabla_h=
 -p_A^\rho\d y^A\wedge\d^{m-2}x_{\mu\rho}+H\d^{m-1}x_\mu
 \\
 \xi_{Y_{\mu\nu}}&=&\inn(Y_{\mu\nu})\Theta^\nabla_h=
 x^\mu(-p_A^\rho\d y^A\wedge\d^{m-2}x_{\nu\rho}+H\d^{m-1}x_\nu)-
 x^\nu(-p_A^\rho\d y^A\wedge\d^{m-2}x_{\mu\rho}+H\d^{m-1}x_\mu)
 \\ &=&
 x^\mu\xi_{Y_\nu}-x^\nu\xi_{Y_\mu}
 \eeann
 If $S\hookrightarrow J^{1*}E$ is an integral submanifold
 of the system, this means that
 $$
 j_S^*\d\xi_{Y_\mu} = 0 \quad , \quad
 j_S^*\d(x^\mu\xi_{Y_\nu}-x^\nu\xi_{Y_\mu}) =
 \d x^\mu\wedge j_S^*\xi_{Y_\nu}-\d x^\nu\wedge j_S^*\xi_{Y_\mu}= 0
 $$

 \section{Conclusions}

 We have used the relation between jet fields (connections) and
 multivector fields in jet bundles to give
 alternative geometric formulations of
 the Hamiltonian equations of first-order Classical Field theories,
 and study their characteristic features.
 In particular:
 \bit
 \item
 The difference between the Hamilton-De Donder-Weyl equations
 and the covariant form of the Hamiltonian equations
 is analyzed and throughly clarified from a geometrical point of view.
 \item
 We prove that the Hamiltonian field equations
 can be written in three equivalent geometric ways:
 using multivector fields in $J^{1*}E$
 (the multimomentum bundle of the Hamiltonian formalism),
 jet fields in $J^1(J^{1*}E)$ or
 their associated Ehresmann connections in $J^{1*}E$. These
 descriptions allow us to write these field equations
 in an analogous way to the dynamical equations for
 (time-dependent) mechanical systems.
 \item
 Using the formalism with multivector fields, we show that
 the field equations $\inn(X_{\cal H})\Omega^\nabla_h=0$,
 with $X_{\cal H}\in\vf^m(J^{1*}E)$ locally decomposable
 and $\bar\tau^1$-transverse,
 have solution everywhere in $J^{1*}E$, which is not
 unique; that is, there are classes of {\sl HDW multivector fields}
 which are solution of these equations. Nevertheless,
 these multivector fields
 are not necessarily integrable everywhere in $J^{1*}E$.
 These features are significant
 differences in relation to the analogous situation in mechanics.
 \item
 The concept of (infinitesimal) symmetry of a Hamiltonian
 system $\hs$ in Field theory is introduced
 and discussed from different points of
 view. The relation between {\sl Cartan-Noether
 symmetries} (those leading to first integrals of
 Noether type) and {\sl general symmetries} has been discussed.
 \item
 In particular,
 a version of Noether's theorem (in the Hamiltonian formalism)
 using multivector fields is proved.
 This statement is also generalized in order to include
 first integrals arising from higher-order Cartan-Noether
 symmetries.
 \item
 We have analyzed the case of
 {\sl restricted Hamiltonian systems}
 (i.e., those such that the Hamiltonian equations
 are stated in a subbundle $P\to E\to M$ of $J^{1*}E$).
 In this case, not even the existence of
 HDW-multivector field is assured, and
 an algorithmic procedure in order to obtain a submanifold of $P$
 where HDW-multivector fields exist, is outlined.
 Of course the solution is not unique, in general.
 \item
 For Hamiltonian systems associated with hyper-regular Lagrangian
 systems in Field theory, we have proved different versions
 of the one-to-one correspondence between the solutions of field
 equations in both formalisms; namely:
 the {\sl equivalence theorem} for sections, jet fields and
 connections, and multivector fields.
 \eit

 Hence, this work completes the results of \cite{EMR-98},
 where the special features of the Lagrangian formalism of
 first-order Field theories in terms of multivector fields
 were studied.

 \appendix
 \section{Appendix}
 \protect\label{mvf}

 (See \cite{EMR-98}, and also \cite{CIL-96b}, \cite{CIL-98} and
 \cite{IEMR-98}).

 Let $E$ be a $n$-dimensional differentiable manifold. Sections of
 $\Lambda^m(\Tan E)$ (with $1\leq m\leq n$) are called
 $m$-{\sl multivector fields} in $E$.
 We will denote by $\vf^m (E)$ the set
 of $m$-multivector fields in $E$.
 Given $Y\in\vf^m(E)$, for
 every $p\in E$, there exists an open neighborhood $U_p\subset E$
 and $Y_1,\ldots ,Y_r\in\vf (U_p)$ such that
 $$
 Y\feble{U_p}\sum_{1\leq i_1<\ldots <i_m\leq r} f^{i_1\ldots
 i_m}Y_{i_1}\wedge\ldots\wedge Y_{i_m}
 $$
 with
 $f^{i_1\ldots i_m}\in\Cinfty (U_p)$ and $m\leq r\leq{\rm dim}\, E$.
 A multivector field
 $Y\in\vf^m(E)$ is {\sl locally decomposable} if, for every $p\in E$,
 there exists an open
 neighborhood $U_p\subset E$ and $Y_1,\ldots ,Y_m\in\vf (U_p)$ such
 that $Y\feble{U_p}Y_1\wedge\ldots\wedge Y_m$.

 If $\Omega\in\df^k(E)$ is a differentiable $k$-form in $E$,
 we can define the contraction
 $$
 \inn(Y)\Omega\feble{U_p} \sum_{1\leq i_1<\ldots <i_m\leq
 r}f^{i_1\ldots i_m} \inn(Y_1\wedge\ldots\wedge Y_m)\Omega =
 \sum_{1\leq i_1<\ldots <i_m\leq r}f^{i_1\ldots i_m} \inn
 (Y_1)\ldots\inn (Y_m)\Omega
 $$
 if $k\geq m$, and equal to zero if $k<m$.
 The $k$-form $\Omega$ is said to be {\sl $j$-nondegenerate}
 (for $1\leq j\leq k-1$) if, for every $p\in E$ and
 $Y\in\vf^j(E)$, $\inn(Y_p)\Omega_p =0\ \Leftrightarrow \ Y_p=0$.
 The graded bracket
 $$
 [\d , \inn (Y)]=\d\inn (Y)-(-1)^m\inn (Y)\d:=\Lie (Y)
 $$
 defines an operation of degree $m-1$ which is called
 the {\sl Lie derivative} respect to $Y$.
 If $Y\in\vf^i(E)$ and $X\in\vf^j(E)$, the graded commutator of
 $\Lie (Y)$ and $\Lie (X)$ is another operation of degree $i+j-2$  of
 the same type, i.e., there will exists a $(i+j-1)$-multivector
 denoted by $[Y,X]$ such that,
 $$
 [\Lie (Y), \Lie (X)] = \Lie([Y,X])
 $$
 The bilinear assignment $X,Y \mapsto [X,Y]$ is called
 the {\sl Schouten-Nijenhuis bracket} of $X,Y$.
 If $X$, $Y$ and $Z$ are multivector fields of degrees $i,j,k$,
 respectively, then the following properties hold:
 \ben
 \item
 $[X,Y] = -(-1)^{(i+1)(j+1)} [Y,X]$.
 \item
 $[X, Y\wedge Z] =
 [X,Y]\wedge Z + (-1)^{(i+1)(j+1)} Y\wedge [X,Z]$.
 \item
 $(-1)^{(i+1)(k+1)}[X,[Y ,Z] ] +
 (-1)^{(j+1)(i+1)}[Y,[Z ,X] ] +(-1)^{(k+1)(j+1)}[Z,[X ,Y] ]
 = 0 $.
 \een
 Moreover, if $X\in\vf^l(E)$ and $Y\in\vf^m(E)$, then
 $$
 \inn([X,Y]) \Omega = \Lie (X)\inn(Y) \Omega -
 (-1)^{l+m}\inn(Y)\Lie (X)\Omega
 $$

 A non-vanishing $m$-multivector field $Y\in\vf^m(E)$
 and a $m$-dimensional distribution $D\subset\Tan E$ are {\sl
 locally associated} if there exists a connected open set
 $U\subseteq E$ such that $Y\vert_U$ is a section of
 $\Lambda^mD\vert_U$. If $Y,Y'\in\vf^m(E)$ are non-vanishing
 multivector fields locally associated with the same distribution
 $D$, on the same connected open set $U$, then there exists a
 non-vanishing function $f\in\Cinfty (U)$ such that
 $Y'\feble{U}fY$. This fact defines an equivalence relation in the
 set of non-vanishing $m$-multivector fields in $E$, whose
 equivalence classes will be denoted by $\{ Y\}_U$. Then, there is
 a bijective correspondence between the set of $m$-dimensional
 orientable distributions $D$ in $\Tan E$ and the set of the
 equivalence classes $\{ Y\}_E$ of non-vanishing, locally
 decomposable $m$-multivector fields in $E$. The distribution
 associated with the class $\{ Y\}_U$ is denoted ${\cal D}_U(Y)$.
 If $U=E$ we write ${\cal D}(Y)$.

A submanifold $S\hookrightarrow E$, with ${\rm dim}\, S=m$, is
said to be an {\rm integral manifold} of $Y\in\vf^m(E)$ if, for
every $p\in S$, $Y_p$ spans $\Lambda^m\Tan_pS$. $Y$ is an {\rm
integrable multivector field} on an open set $U\subseteq E$ if,
for every $p\in U$, there exists an integral manifold
$S\hookrightarrow U$ of $Y$, with $p\in S$. $Y$ is {\rm
integrable} if it is integrable in $E$. $Y$ is {\rm involutive} on
a connected open set $U\subseteq E$ if it is locally decomposable
in $U$ and its associated distribution ${\cal D}_U(Y)$ is
involutive. $Y$ is {\rm involutive} if it is involutive on $E$. If
a $Y\in\vf^m(E)$ is integrable, then so is every other in its
equivalence class $\{ Y\}$, and all of them have the same integral
manifolds. Moreover, {\sl Frobenius' theorem} allows us to say
that a non-vanishing and locally decomposable multivector field is
integrable on a connected open set $U\subseteq E$ if, and only if,
it is involutive on $U$.

Now, let $\pi\colon E\to M$ be a fiber bundle. $Y\in\vf^m(E)$ is
said to be {\sl $\pi$-transverse} if, at every point $y\in E$,
$(\inn (Y)(\pi^*\omega))_y\not= 0$, for every $\omega\in\df^m(M)$
with $\omega (\pi(y))\not= 0$. Then, if $Y\in\vf^m(E)$ is
integrable, $Y$ is $\pi$-transverse if, and only if, its integral
manifolds are local sections of $\pi\colon E\to M$. In this case,
if $\phi\colon U\subset M\to E$ is a local section with $\phi
(x)=y$ and $\phi (U)$ is the integral manifold of $Y$ through $y$,
then $\Tan_y({\rm Im}\,\phi)$ is ${\cal D}_y(Y)$.

 In Hamiltonian Field theory we are interested in multivector
 fields in $\bar\tau^1\colon J^{1*}E\to M$.
 Now remember that a {\sl connection} in $J^{1*}E$ is one of the
 following equivalent elements:
 a global section ${\cal Y}\colon J^{1*}E\to J^1(J^{1*}E)$
 of the projection $J^1(J^{1*}E)\to J^{1*}E$ (a {\sl jet field}),
 a subbundle ${\rm H}(J^{1*}E)$ of $\Tan J^{1*}E$ such that
 $\Tan J^{1*}E={\rm V}(\bar\tau^1)\oplus{\rm H}(J^{1*}E)$
 (which is called a {\sl horizontal subbundle}, and
 it is also denoted by ${\cal D}({\cal Y})$ when considered as the
 distribution associated with ${\cal Y}$),
 or a $\bar\tau^1$-semibasic $1$-form $\nabla$ on $J^{1*}E$
 with values in $\Tan J^{1*}E$, such that
 $\nabla^*\alpha =\alpha$, for every $\bar\tau^1$-semibasic
 form $\alpha\in\df^1(J^{1*}E)$
 (the {\sl connection form} or {\sl Ehresmann connection}).
 A jet field ${\cal Y}\colon J^{1*}E\to J^1(J^{1*}E)$
 (or a connection $\nabla$) is {\sl orientable} if
 ${\cal D}({\cal Y})$ is an orientable distribution on $J^{1*}E$. Then:

 \begin{teor} \quad
 There is a bijective correspondence between the set of orientable
 jet fields ${\cal Y}\colon J^{1*}E\to J^1(J^{1*}E)$ (or orientable
 connections $\nabla$ in $\bar\tau^1\colon J^{1*}E\to M$) and the set
 of the equivalence classes of locally decomposable and
 $\bar\tau^1$-transverse multivector fields
 $\{X\}\subset\vf^m(J^{1*}E)$ (they are characterized by the fact that
 ${\cal D}({\cal Y})={\cal D}(X)$).
 Then, ${\cal Y}$ is integrable,
 if, and only if, so is $X$, for every $X\in\{ X\}$.
 \label{vfmvf}
 \end{teor}

 The expression for a representative
 multivector field $X$ of the class $\{ X\}$ associated with a jet field
 ${\cal Y}\equiv
 (x^\mu,y^A,p_A^\mu,F_\mu^A(x,y,p),G_{A\mu}^\rho(x,y,p))$
 is
 \dst X=\bigwedge_{\mu=1}^m
 \left(\derpar{}{x^\mu}+F_\mu^A\derpar{}{y^A}+
 G_{A\mu}^\rho\derpar{}{p^\rho_A}\right) \) .

 \subsection*{Acknowledgments}

 We wish to thank Prof. M. Fern\'andez-Ra\~nada for clarifying
 us some questions about symmetries of mechanical systems.
 We also thank Mr. Jeff Palmer for his
 assistance in preparing the English version of the manuscript.
 We are grateful for the financial support of the CICYT
 TAP97-0969-C03-01.

 \end{document}